\documentclass[draftcls, 12pt, onecolumn]{IEEEtran}
\normalsize
\usepackage{amsfonts}
\usepackage{times}
\usepackage[pdftex]{graphicx}
\usepackage{latexsym}
\usepackage{dsfont}
\usepackage{amssymb}
\usepackage{amsmath}
\usepackage{cite}
\usepackage{verbatim}
\usepackage{subfigure}

\def\bb0{{\mathbb{0}}}

\def\bb{{\mathbf{b}}}

\def\b0{{\mathbf{0}}}

\def\bH{{\mathbf{H}}}

\def\bU{{\mathbf{U}}}

\def\bX{{\mathbf{X}}}

\def\sf0{{\mathsf{0}}}

\usepackage{color,graphicx,epsfig,psfrag}
\usepackage{pifont,enumerate,cases,algorithm,balance}
\usepackage{mathrsfs} 
\usepackage{multirow,multicol}
\usepackage[table]{xcolor}

\def\r0{{\mathbf{0}}}

\def\rA{{\mathbf{A}}}
\def\rB{{\mathbf{B}}}

\def\rG{{\mathbf{G}}}
\def\rH{{\mathbf{H}}}

\def\rP{{\mathbf{P}}}

\def\rU{{\mathbf{U}}}

\def\rW{{\mathbf{W}}}
\def\rX{{\mathbf{X}}}

\def\rZ{{\mathbf{Z}}}

\def\b0{\mathbf 0}
\def\b1{\mathbf 1}

\def\r{\text{r}}

\def\vec{\mathrm{vec}}

\usepackage{bm} 

\usepackage{afterpage}
\usepackage{subfigure}

\usepackage{algorithm}
\usepackage{algpseudocode}

\usepackage{mathtools}
\usepackage{bbm}
\usepackage{url}
\usepackage{setspace}

\DeclarePairedDelimiter\floor{\lfloor}{\rfloor}

\usepackage{amsmath}
\usepackage{graphicx}

\makeatletter
\newcommand{\ostar}{\mathbin{\mathpalette\make@circled\star}}
\newcommand{\make@circled}[2]{%
  \ooalign{$\m@th#1\smallbigcirc{#1}$\cr\hidewidth$\m@th#1#2$\hidewidth\cr}%
}
\newcommand{\smallbigcirc}[1]{%
  \vcenter{\hbox{\scalebox{0.77778}{$\m@th#1\bigcirc$}}}%
}
\makeatother

\title{Deep Learning-based Compressive Beam Alignment in mmWave Vehicular Systems}
\author{\IEEEauthorblockN{Yuyang Wang, Nitin Jonathan Myers, Nuria Gonz\'alez-Prelcic, \\
and Robert W. Heath Jr. }\thanks{Yuyang Wang is with Apple Inc., One Apple park way, Cupertino, CA, 95014, USA, email: yuywang@utexas.edu. Nitin Jonathan Myers is with Samsung Semiconductor Inc., 5465 Morehouse Dr, San Diego, CA 92121 USA, email: nitinjmyers@utexas.edu. Nuria Gonz\'alez-Prelcic, and Robert W. Heath Jr. are with the Department of Electrical and Computer Engineering, North Carolina State University, 
890 Oval Dr, Raleigh, NC 27606 USA, email:  $\{$ngprelcic, rwheathjr$\}$@ncsu.edu. Part of this work has been presented at IEEE ICASSP 2020 \cite{myers2020deep}.
This material is based upon work supported in part by the National Science Foundation under Grant No. ECCS-1711702, and by a Qualcomm Faculty Award. Work was performed when Y. Wang and N. J. Myers were at UT Austin.}}

\begin{document}
\maketitle
\begin{abstract}
Millimeter wave vehicular channels exhibit structure that can be exploited for beam alignment
with fewer channel measurements compared to exhaustive beam search. With fixed layouts of roadside
buildings and regular vehicular moving trajectory, the dominant path directions of channels will likely
be among a subset of beam directions instead of distributing randomly over the whole beamspace. In this
paper, we propose a deep learning-based technique to design a structured compressed sensing (CS) matrix
that is well suited to the underlying channel distribution for mmWave vehicular beam alignment. The
proposed approach leverages both sparsity and the particular spatial structure that appears in vehicular
channels. We model the compressive channel acquisition by a two-dimensional (2D) convolutional
layer followed by dropout. We design fully-connected layers to optimize channel acquisition and beam
alignment. We incorporate the low-resolution phase shifter constraint during neural network training by
using projected gradient descent for weight updates. Furthermore, we exploit channel spectral structure
to optimize the power allocated for different subcarriers. Simulations indicate that our deep learningbased approach achieves better beam alignment than standard CS techniques which use random phase
shift-based design. Numerical experiments also show that one single subcarrier is sufficient to provide
necessary information for beam alignment.
\end{abstract}
\section{Introduction}
\par Millimeter-wave (mmWave) vehicular communication enables massive sensor data sharing and various emerging applications related to safety, traffic efficiency and infotainment \cite{wang2018mmwave, va2016millimeter, giordani2017millimeter}. Beam alignment in vehicular communication, however, is very challenging due to the use of large antenna arrays and high mobility of vehicles \cite{heathoverview, choi2016millimeter, wang2018mmwave}. Beams at the transceivers can be selected based on multi-path signatures measured through beam training. The exhaustive search-based beam training can result in a substantial overhead at mmWave. Such overhead fails to satisfy the low latency requirements in vehicular applications due to highly dynamic channels and the short beam coherence time
 \cite{11ad}. 
 
 \par A possible approach to reduce this overhead is to exploit sparsity of mmWave channel and leverage compressed sensing (CS)-based methods that acquire a lower dimensional channel representation \cite{csintro}. The sparse nature of mmWave channels in an appropriate dictionary makes CS  a promising solution for mmWave channel estimation or beam alignment with few channel measurements. Channel recovery-based on random compressive channel projections is able to achieve fast and accurate beam alignment \cite{cschest}. In typical vehicular communication scenarios, channels may exhibit structure in the space of beam directions beyond sparsity \cite{yuyang_access, wang2020sitespecific, myers2020deep}. Designing random phase-shift-based CS that exploits underlying channel prior is a challenging task. In this paper, we investigate convolutional compressive sensing (CCS)-based solution, where the signal is projected onto circulant shift of a known signal  \cite{convCS}.  We examine 2D-CCS, the extension of CCS to planar arrays, where the base station (BS) applies 2D-circulant shifts of a base matrix to its antenna array for measuring the channel \cite{falp}. The advantages of 2D-CCS are two-fold: 1) it can be easily parameterized by a single base matrix, and 2) proper design of the base matrix can exploit the statistical pattern in the channel to acquire channel measurements more efficiently. In \cite{wang2020sitespecific}, we observed that the statistical pattern of channels in a site-specific vehicualr scenario is closely related to the surrounding street layouts and the vehicle moving trajectory. We showed that the phase shift matrix can be optimized in an online manner based on the available angular domain prior cached at the BS. In this paper, we will demonstrate how deep learning can be leveraged to optimize, not only the phase shift matrix, but also the channel recovery and beam alignment components in CS which were theoretically derived and optimized in \cite{wang2020sitespecific}. 
\par Deep learning can be leveraged to exploit structure of the signals for various tasks in CS \cite{khobahi2019deep, he2018deep, mousavi2017deepcodec}. For example, deep learning can be designed for sensing and sparse signal recovery with significantly reduced complexity and running time \cite{mousavi2015deep, mousavi2017learning, palangi2016distributed, convcsnet}. Data-driven solutions make it possible to learn a representation of signals from a set of representative signals, and enables recovery of the original signal from fewer number of measurements. Furthermore, it was shown that deep learning can be an alternative to CS-based mmWave channel estimation \cite{guo2020convolutional, li2019deep, wang2018deep}. For example, in \cite{li2019deep}, an auto-encoder type one-dimensional (1D) convolutional neural network (CNN) was trained to design the hybrid precoding matrices using a few training pilots. The CNN in \cite{li2019deep} was trained using information about the surrounding environment and user distributions. The beamforming vectors were obtained from the optimized parameters in the neural network. In our paper, we consider a similar autoencoder-type architecture that achieves CS design using a CNN with a two-dimensional (2D) convolutional layer. Instead of using a vectorized channel as an input, we exploit the 2D channel structures that can be captured by the convolutional layer. We build the CS-based beam alignment solution on a 2D-convolutional CS (2D-CCS) framework, which is able to provide a high level of interpretability in the results when compared to the approach in \cite{li2019deep}. We design the 2D-convolutional layer to emulate the channel acquisition and subsampling in 2D-CCS instead of the Kronecker product in \cite{li2019deep}. 
\par 
In this paper, we propose a structured CS matrix optimization framework for mmWave vehicular beam alignment. We consider an urban mmWave vehicle-to-infrastructure (V2I) scenario where a roadside unit (RSU) is deployed on the road side to provide coverage to the vehicles passing by. The model leverages previous transmission records at the BS to design a proper phase shift matrix for beam alignment. We do not assume the availability of any side information such as various types of sensor data, while the model optimizes a general site-specific phase shift matrix that is well-suited to the underlying channel statistics. Furthermore, the deep learning-based model considers  phase shift matrix design with low-resolution phased array hardware constraints. The main contributions are summarized as follows.
\begin{itemize}
\item We propose a novel deep learning-based CS optimization solution for mmWave vehicular beam alignment. We propose a 2D-CCS framework for CS and beam alignment. We show that complex CNN can be leveraged to replace the different components of 2D-CCS, optimize the phase shift matrix and predict optimal beam configuration for mmWave vehicular communication. In particular, the convolutional layers are used to acquire the compressive channel measurements, and the fully-connected layers are leveraged to conduct beam alignment. 

\item We optimize the BS phase shift matrix design with low-resolution phase shift hardware constraint. We propose projected gradient descent-based approach to incorporate phase quantization in the weight update during training. We show that projected gradient descent that is implemented inside the training can optimize a CS matrix that is subject to low-resolution phase shift constraint. We show that proper preprocessing of the channel for the training and testing data can improve the model's robustness to noise and enhance the prediction accuracy. Furthermore, our model exhibits good interpretability in the results. We show that the optimized phase shift matrix is well-suited to the site-specific channel statistics related to the spatial distribution of roads. 

 \item We extend the CS framework optimization to wideband channels. The wideband model shares a similar CNN framework as the narrowband one, but takes a concatenation of different channel subcarriers as input. We propose to formulate a linear layer after the 2D convolutional layer to optimize the power allocated to different subcarriers in the input. We apply special constraints towards the linear layer weights to impose scaling optimization of different subcarriers.   \end{itemize}
 
 The rest of the paper is organized as follows. We motivate the problem and explain the specific structure in the vehicular channels that is related to the site-specific street layouts in Section \ref{sec:pattern}. We provide the simulation setup and channel model in Section \ref{sec:simulationsetup2} and \ref{sec:channelmodel}. In Section \ref{sec:2dccs} and \ref{sec:2dccs_implem}, we explain the basics of 2D-CCS and demonstrate how to use deep learning to mimic the end-to-end 2D-CCS framework in a narrowband system. Section \ref{sec:wideband} shows the extension of the proposed framework to wideband systems. Comprehensive numerical results are demonstrated in Section \ref{sec:simulationlast}. The final conclusions are drawn in Section \ref{sec:conclusionlast}. 

\par  \textbf{Notation}$:$ $\mathbf{A}$ is a matrix, $\mathbf{a}$ is a column vector and $a, A$ denote scalars. $\rA^T$, $\overline{\rA}$ and $\rA^*$ represent the transpose, conjugate and conjugate transpose of A.The real and imaginary parts of $\mathbf{A}$ are denoted by $\mathbf{A}_{\mathrm{R}}$ and $\mathbf{A}_{\mathrm{I}}$. $A(k,\ell)$ denotes the entry of $\mathbf{A}$ in the $k^{\mathrm{th}}$ row and the ${\ell}^{\mathrm{th}}$ column. The $\ell^{\mathrm{th}}$ column of $\mathbf{A}$ is denoted by $\mathbf{A}(:,\ell)$. $|\mathbf{A}|$ is a matrix that contains the magnitude of the entries in $\mathbf{A}$. The Frobenius norm of $\mathbf{A}$ is $\Vert \mathbf{A} \Vert_{\mathrm{F}}$. The inner product of two matrices $\mathbf{A}$ and $\mathbf{B}$ is defined as $\langle \mathbf{A},\mathbf{B}\rangle =\sum_{k,\ell}A(k,\ell)\overline{B}(k,\ell)$. The multiplication of two matrices $\rA$ and $\rB$ is represented as $\rA\rB$. The matrix $[\mathbf{A};\mathbf{B}]$ is obtained by vertically stacking $\mathbf{A}$ and $\mathbf{B}$. $\mathbf{U}_N$ is an $N\times N$ unitary discrete Fourier transform (DFT) matrix. $\mathsf{j}=\sqrt{-1}$.  
\section{Motivation and system model}\label{sec:motivaion}
In this section, we illustrate the motivation behind the deep learning-based phase shift matrix design and beam alignment in mmWave vehicular communication. We explain the system model, the data collection and preprocessing procedures to establish the channel dataset. We demonstrate how the specific structure in vehicular urban layouts can lead to statistical pattern in the channel and benefit a more efficient CS design.
\subsection{Statistical channel patterns}\label{sec:pattern}
There exists statistical patterns in mmWave vehicular channels that are related to the street layouts. In a typical vehicular scenario, for example, an urban canyon, a number of objects are stationary such as buildings, roads and roadside infrastructures. With fixed moving trajectories of vehicles, the angular distribution of the channel paths follows a statistical non-uniform distribution corresponding to the moving trajectory and surrounding environment. The availability of an informative angular prior can be leveraged for efficient beam alignment and channel estimation in mmWave vehicular communication. For example, weighted sparse recovery techniques that exploit such a prior can be used instead of standard CS algorithms for mmWave link configuration [14]. Furthermore, the beam training vectors used to acquire channel measurements can be optimized based on the channel prior to increase the probability of successful alignment. 
\subsection{Simulation setup}\label{sec:simulationsetup2}
\begin{figure}[!htb]
    \centering
    \begin{minipage}{.45\textwidth}
        \centering
        \includegraphics[width=3.0in]{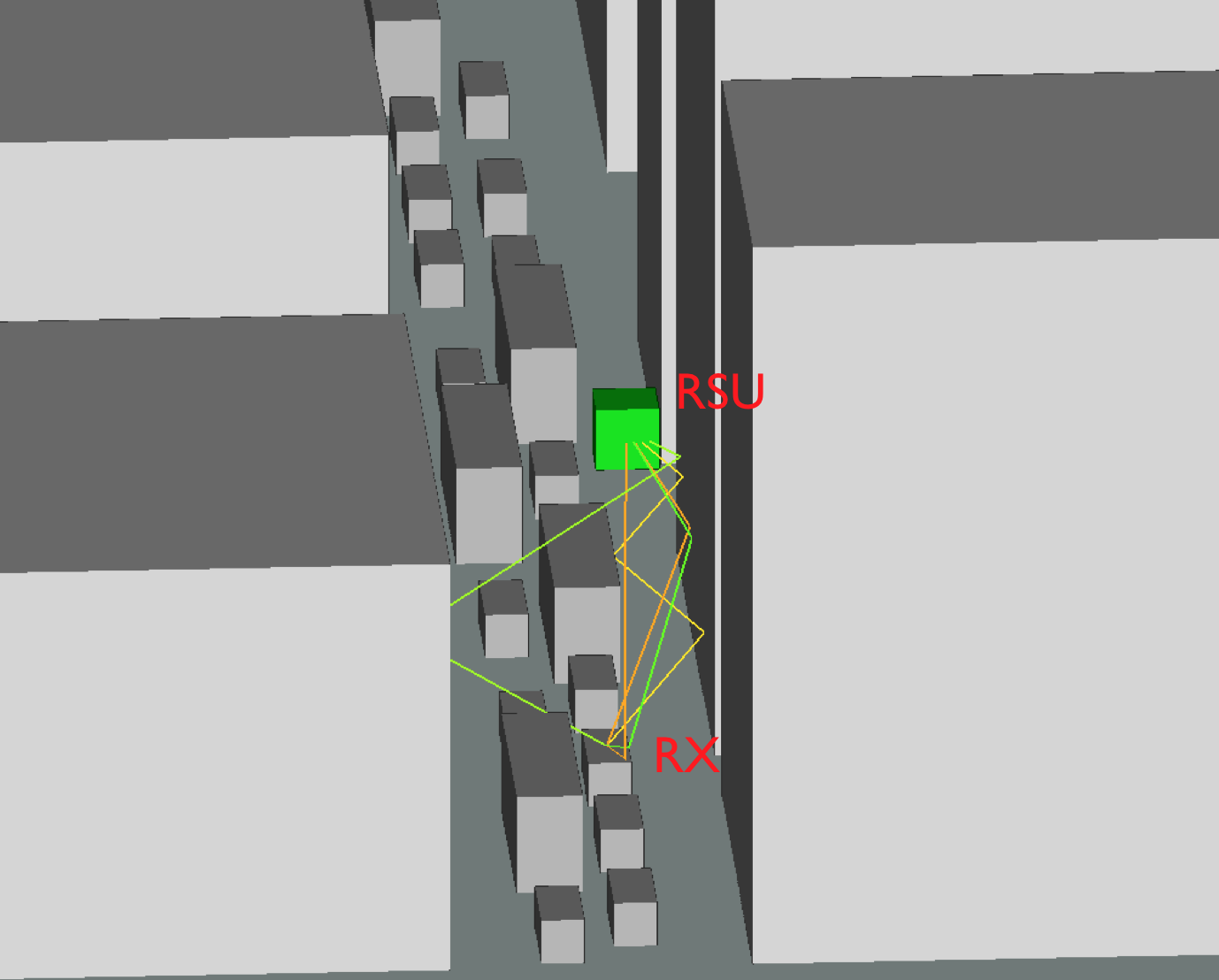}
\caption{Illustration of the ray tracing setup in which cars and trucks are randomly dropped in the two lanes of the urban canyon. Receivers are mounted on the top center of the low vehicle. The BS is mounted on a street-side lamp-post. The figure illustrates the \emph{top five} strongest paths of the channel for a certain receiver. Our channel model includes the effect of multiple reflections that occur at the buildings and the vehicles. }\label{fig:simulationsetup}
    \end{minipage}%
\hfill
    \begin{minipage}{0.45\textwidth}
        \centering
        \includegraphics[trim = 6cm 0cm 0cm 1cm, clip = true, width=3.5in]{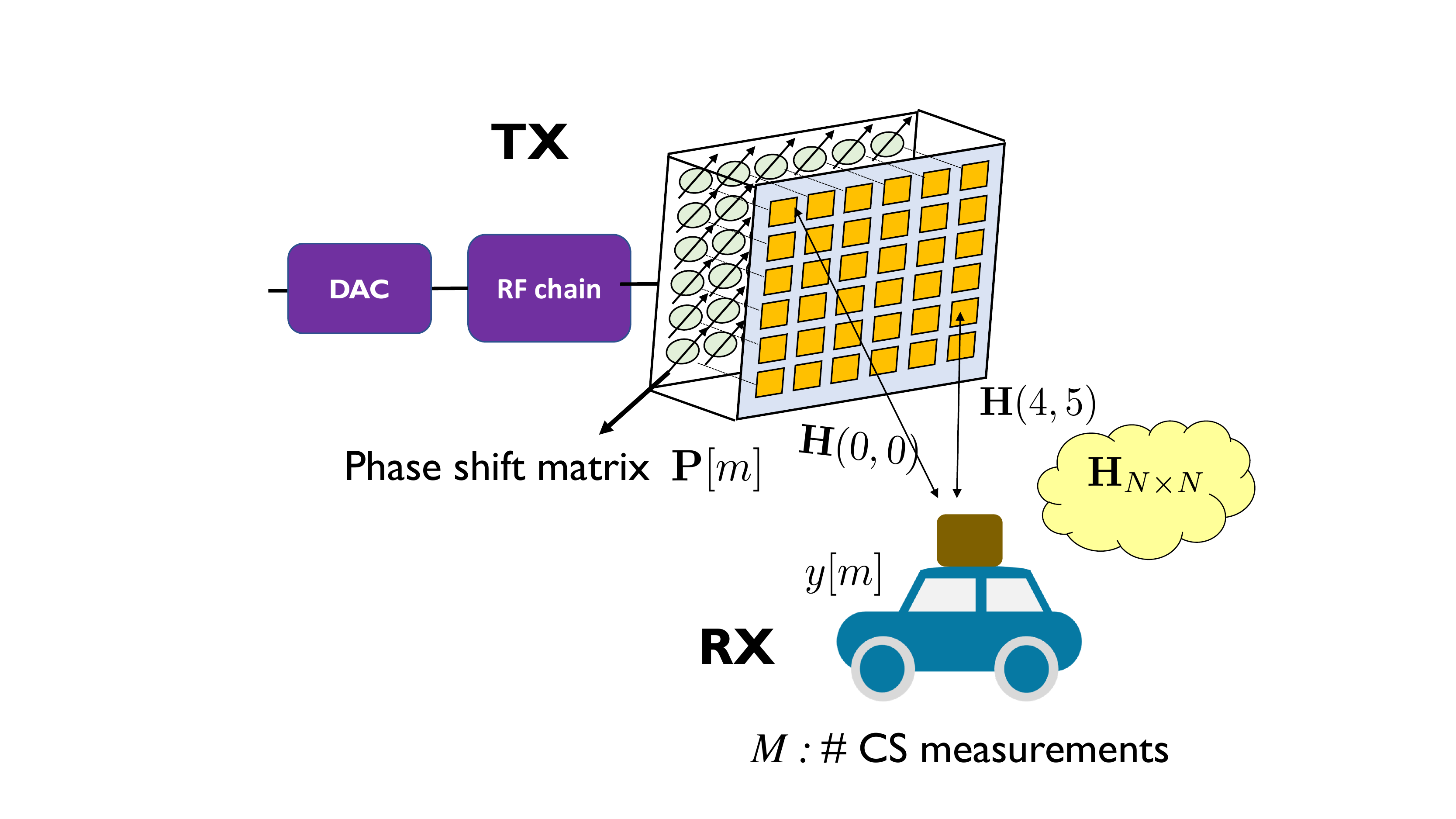}
\caption{An illustration of the system model in 2D-CCS. The BS has a single RF chain and is equipped with a uniform planar phased array of size $N\times N$, and there is one single antenna at the receiver. The BS applies phase shift matrix in successive $M$ time slots. The $m$-th CS channel measurement is the projection of the channel on the phase shift matrix $\rP[m]$.}\label{fig:systemmodel}
    \end{minipage}
\end{figure}

In this paper, we consider a mmWave V2I scenario in an urban canyon. Since there is no widely available mmWave V2I testbed yet, we use ray tracing technique to simulate the vehicular channel. We use Wireless Insite, a commercial ray tracing simulator, to establish the channel dataset for evaluation \cite{wireless_insite}. Ray tracing simulation projects rays from the BS to the physical environment and calculates the channel path information, such as path gain, angle-of-arrival (AoA) and angle-of-departure (AoD). An example of the ray tracing setup is shown in Fig. \ref{fig:simulationsetup}. We model buildings as cuboids with a concrete exterior, located on the two road sides with different sizes. For simplicity, we consider two types of vehicles: trucks and sedans. We model the vehicles as cuboids with metal exteriors, where types of vehicles are differentiated by the sizes. The vehicles are randomly dropped on the two lanes. More details of the ray tracing setup can be found in \cite{yuyang_access}. The antenna panels at the receiver are placed at the vehicle rooftop, facing the sky, the antenna panel at the BS faces towards the street. In our simulation, we obtain a total of $L$ output paths for each channel realization, which includes the AoD azimuth $\phi^\mathrm{D}_\ell$, elevation $\theta^{\mathrm D}_\ell$, the AoA azimuth $\phi^\mathrm{A}_\ell$, elevation $\theta^{\mathrm A}_\ell$, the time-of-arrival $\tau_\ell$, the path gain $\alpha_\ell$, and the phase $\beta_\ell$. The output of the ray tracing simulation is used to calculate the vehicular channels. 

\subsection{Narrowband channel modeling}\label{sec:channelmodel}
An illustration of the system model is provided in Fig. \ref{fig:systemmodel}. The BS is equipped with a square uniform planar array (UPA) of size $N\times N$. We consider an analog beamforming architecture at the BS. In such an architecture, every antenna of the UPA is connected to a single radio frequency chain through a phase shifter. The use of phase shifters allows the BS to generate a variety of beams that can be used for initial access or data transmission. As the focus of this work is on the transmit beam alignment problem, we assume a single antenna receiver at the vehicles.  Under the multiple-input single output (MISO) system assumption, the antenna domain channel can be represented as a vector $\mathbf{h} \in \mathbb{C}^{N^2 \times 1}$. The channel vector $\mathbf{h}$ can be reshaped into a matrix $\mathbf{H}\in\mathbb{C}^{N\times N}$, with $ \mathbf{h} = \vec{(\rH)}$. The $(r, c)^\mathrm{th}$ entry of $\rH$ represents the channel coefficient between the $(r, c)^\mathrm{th}$ antenna at the transmitter and the receiver antenna. We assume a half-wavelength spaced UPA at the BS to define the Vandermonde vector $\mathbf{a}_N(\Delta)$ as 
\begin{align}
\mathbf{a}_N(\Delta) =  [1, ~e^{\mathsf{j}\pi\Delta}, ~e^{2\mathsf{j}\pi\Delta}, ~\cdots, ~e^{(N - 1)\mathsf{j}\pi\Delta}]. 
\end{align}

For tractability of analysis and simplicity of explanation, we assume that the channel is narrowband. The channel between the BS and the receiver can be expressed as 
\begin{align}
\label{eq:channel_expansion}
\mathbf{H}= \sum_{\ell = 1}^{L_\mathrm{p}} \alpha_\ell e^{\mathsf{j}\beta_\ell}\mathbf{a}_N(\cos\theta_\ell)\mathbf{a}_N{(\sin\theta_\ell\cos\phi_\ell)}^T. 
\end{align}
 The matrix representation of the MISO channel in (\ref{eq:channel_expansion}) allows us to better explain the ideas underlying our beam training design. In the channel calculation, we consider both line-of-sight (LOS) and non line-of-sight (NLOS) paths in the channel. 
  
 We consider a 2D-CCS-based beam alignment framework. The phased array at the BS has low-resolution phase shifters. Denoting the number of bits as $q$, we use $\mathbb{Q}_q$ to represent the set of possible antenna weights in the analog beamforming network, i.e., $\mathbb{Q}_q=\{e^{\mathsf{j} 2 \pi b/ 2^q}/ N : b \in \{1,2,\cdots 2^q\}\}$. As illustrated in Fig. \ref{fig:systemmodel}, the BS applies different phase shift matrices to its antenna array in a total of $M$ successive training slots. Since we assume an omnidirectional antenna at the receiver, the set of the measurement is simply the projections of the channel on the phase shift matrices used by the BS in different training slots. In the $m$-th measurement slot, $1\leq m \leq M$, the phase shift matrix is represented as $\rP[m] \in \mathbb{C}^{N\times N}$.  Since $\mathbf{P}[m]$ is the phase shift matrix generated by the phased array, it is subject to unit-norm constraint per phase shifter. Defining $\mathbf{1}$ as an all-one matrix of size $N\times N$, we have $\left|\mathbf{P}[m]\right| = \mathbf{1}$, and $\Vert {\mathbf{P}}[m] \Vert_{\mathrm F}=N, ~\forall m \in \{1, 2, \cdots, M\}$.
  We denote the additive white Gaussian noise (AWGN) as $v[m] \in \mathbb{C}$. The received signal can be written as  
 \begin{align}
 \label{eq:sysmodel}
 y[m] = \langle \mathbf{H}, \mathbf{P}[m]\rangle + v[m], ~ m = 1, 2, \cdots, M.
 \end{align}
 Finally, the receiver feedbacks the channel measurement vector $\{{y}[m]\}_{m = 1}^M\in\mathbb{C}^{N\times1}$ to the BS using the control channel, for channel estimation and  the subsequent beam alignment.

 We define $\mathcal{F}$ as the standard 2D-DFT codebook for the UPA-based transmitter. Based on the sequence of channel measurement feedback $\{{y}[m]\}_{m = 1}^M\in\mathbb{C}^{N\times1}$, the BS applies CS techniques to estimate the channel so that the BS can select the beamformer  $\mathbf{P}_{\mathrm{BF}} \in \mathcal{F}$ that has a maximum inner product with the channel estimate $|\langle \mathbf{\hat{H}},\mathbf{P}\rangle|$ to align the beam. In this paper, we focus on solving two challenging problems: 1) phase shift matrix optimization, i.e., $\{\mathbf{P}[m]\}_{m = 1}^M$, that leverages available channel prior for compressive measurement acquisition, and 2) prediction of optimal beamformer $\mathbf{P}_{\mathrm{BF}}$ that maximizes the beamforming gain. 

  \par
  Channels exhibit approximate sparsity at mmWave carrier frequencies in the 2D-DFT dictionary \cite{heathoverview} due to the less favorable propagation characteristics at higher frequency. As illustrated in Fig. \ref{fig:systemmodel}, the antenna at the receiver is omnidirectional, and we use a matrix $\bH\in\mathbb{C}^{N\times N}$ to denote the channel represented by the UPA. The 2D-DFT beamspace of the channel can be defined as the 2D-DFT of $\bH$, i.e., 
  \begin{align}\label{equ:beamspace2D}
  \bX = \bU_N^*\bH\bU_N^*. 
  \end{align}
  The beamspace contains the \emph{virtual} channel coefficients seen when different 2D-DFT-based beams are used at the BS \cite{brady2013beamspace}. The support of the non-zero coefficients in $\rX$ indicates the distribution of AoDs of the propagation rays in the channel. Thanks to the limited scattering characteristics at mmWave frequencies, the number of propagation paths in the channels is much less compared to the rich scattering in lower frequencies, which makes the channel beamspace approximately sparse. 
  
A straightforward approach for beam alignment is to either have an estimate of all the $N^2$ coefficients of $\mathbf{H}$ using scalar projections of the form in \eqref{eq:sysmodel} or the 2D-DFT beamspace defined in (\ref{equ:beamspace2D}), followed by beam alignment based on the estimated channel.  Estimating the $N \times N$ channel matrix or the beamspace, however, can result in a significant overhead as the channel dimension is large in typical mmWave settings. Fortunately, sparsity of the channel beamspace $\rX$ allows the use of CS-based algorithms for fast channel estimation or beam alignment. Prior work has shown that beam alignment can be performed with just $\mathcal{O}(\mathrm{log}\, N)$ random phase shift-based compressed measurements of a sparse channel \cite{falp}. It may be possible to estimate the best beam without explicit channel estimation and large overhead. Furthermore, special cases like vehicular channels exhibit certain channel structure due to the fixed street layouts in the surroundings \cite{myers2020deep}, \cite{wang2020sitespecific}. The random phase shift-based CS may perform well. Is it possible, though, to construct a CS matrix that is better matched to the channel prior in vehicular settings that enables a more efficient compressive channel acquisition?
  \begin{figure}[htbp]
  	\vspace{-6mm}
  	\centering
  	{\includegraphics[trim=0cm 0cm 1.5cm 0.5cm,clip=true, width=7.0cm, height=6.5cm]{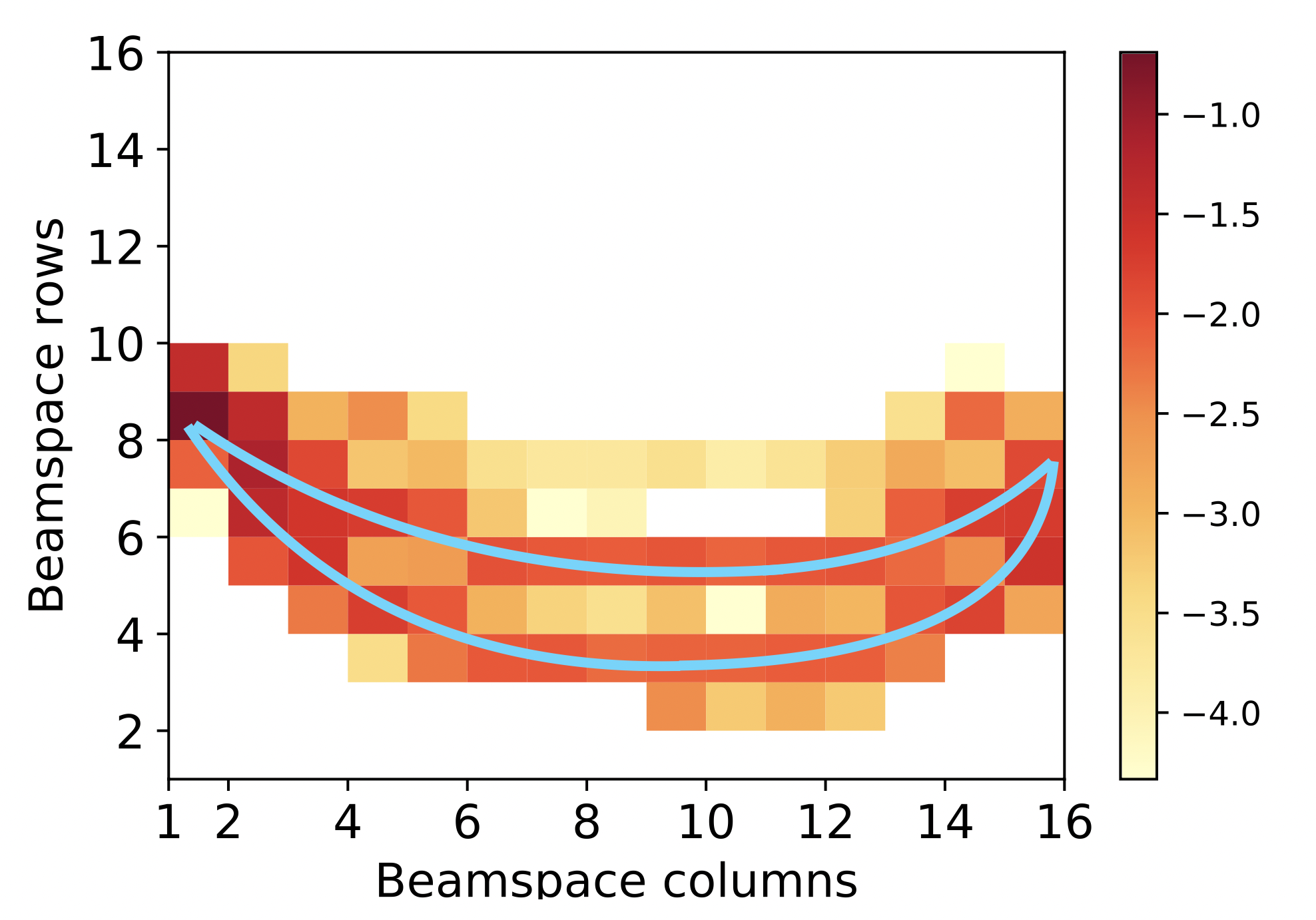}}
  	{\includegraphics[trim=0cm 0cm 1.5cm 1cm,clip=true, width=7.0cm, height=6.5cm]{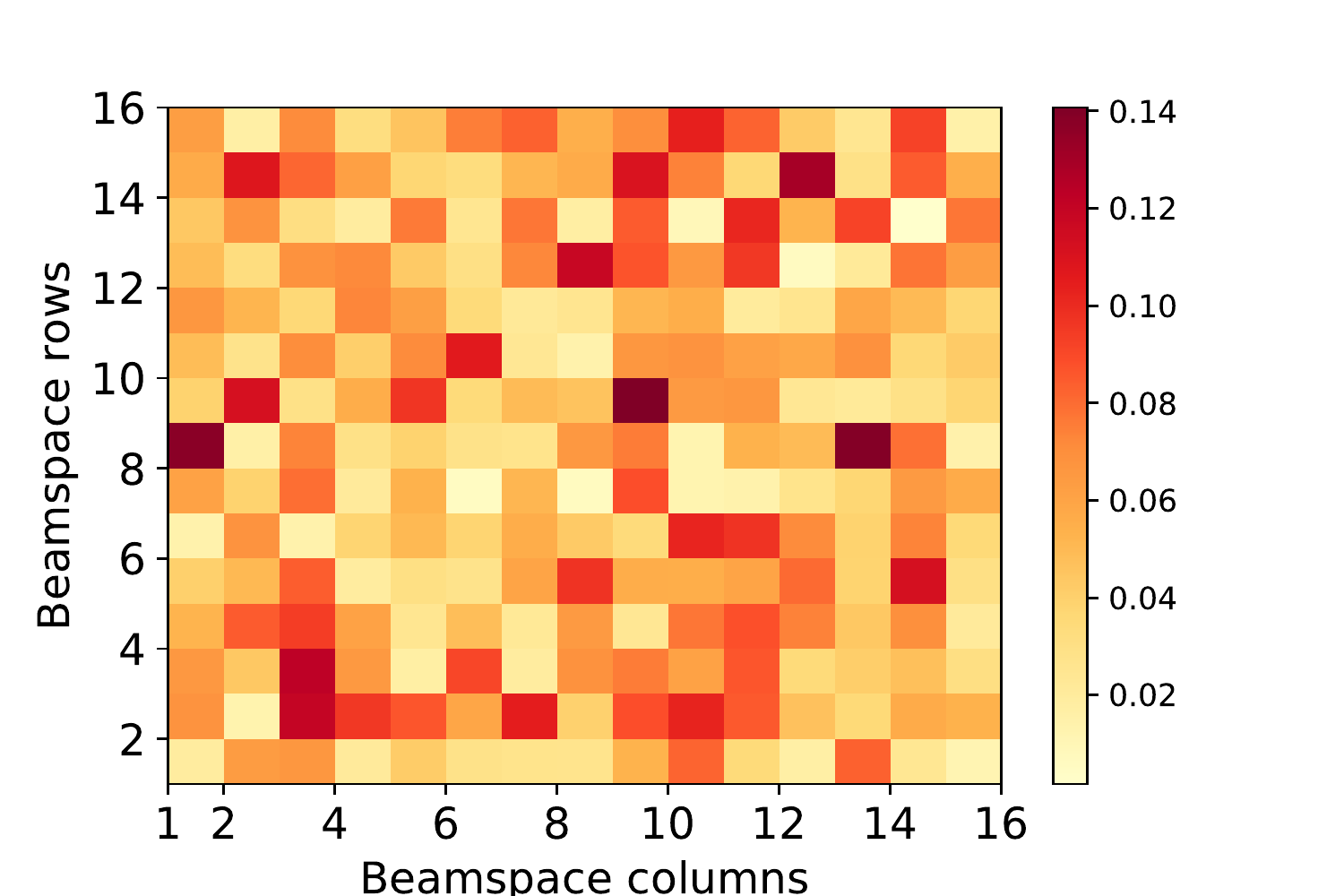}}
  	\vspace{1mm}
  	\caption{ \small For the vehicular communication scenario in our simulations, the beamspace prior is non-zero over a small set in the space of 2D-DFT directions. The blue curves represent the LoS directions associated with vehicles on the two lanes. Random phase shift-based CS uses quasi-omnidirectional beams and does not exploit such an information.  
  		\normalsize}
  	\label{fig:beamspace_prior}
  	\vspace{-2mm}
  \end{figure}
  \par We now describe the additional channel structure present in a typical mmWave vehicular communication scenario, as illustrated in Fig.~\ref{fig:simulationsetup}. For an ensemble of receivers within the coverage of an RSU, we show the probability distribution of the best transmit beams in Fig. \ref{fig:beamspace_prior}, which quantifies the occurrences as the beam being the strongest in the 2D-DFT beamspace. In particular, the $(i,j)^{\mathrm{th}}$ entry of the beamspace prior in Fig. \ref{fig:beamspace_prior} is the probability that the $(i,j)^{\mathrm{th}}$ 2D-DFT beam, i.e., $\mathbf{U}_N(:,i)\mathbf{U}^T_N(:,j)$, provides the largest beam received power for the receiving vehicle. In our simulation scenario, vehicles move on a two-lane straight street, which leads to the existence of two strips along the azimuth direction on an $N \times N$ grid of 2D-DFT angles. These two strips correspond to the ensemble of channels with the strongest beams pointing towards the vehicle moving trajectories associated with two different lanes shown in Fig. \ref{fig:simulationsetup}, i.e., the LOS channels free of blockage. Other beam directions that have the occurrences of being optimal correspond to the channels subject to blockages. It should be noted that the beamspace probability is generally concentrated in a small support set, with a lot of beam directions bearing a very slight chance of being optimal. Compressed sensing using random phase shift matrices, however, does not exploit such specific channel statistics. This is because the beam associated with a random phase shift matrix is quasi-omnidirectional with a high probability; an example of such a beam is shown in Fig. \ref{fig:beamspace_prior}. Sensing with a quasi-omnidirectional pattern is inefficient in vehicular applications, since the vehicle locations are constrained to the road lanes, and many angular directions can be discarded apriori.
  
\section{Deep learning-based 2D-CCS optimization}\label{sec:beamalign}
In this section, we propose a deep learning-based CS framework optimization for mmWave vehicular beam alignment. We introduce 2D-CCS for beam alignment and show how deep neural networks can be tailored to optimize dfiferent parameters in 2D-CCS. We mainly focus on narrowband channels in this section; we extend to wideband channels in Section \ref{sec:wideband}. 

\subsection{2D-CCS: advantages over standard CS}\label{sec:2dccs}
\par 
In 2D-CCS, the BS applies distinct \emph{2D circulant shifts} of the a \emph{base} phase shift matrix $\mathbf{P}$ to its phased array for the receiver to acquire channel measurements in $M$ successive time slots \cite{falp}. We define the row shift sequence in the $M$ training slots as $\{r[m]\}_{m = 1}^M$ and the column shift sequence as $\{c[m]\}_{m = 1}^M$. In the $m$-th time slot, the BS applies phase shift matrix $\rP[m]$ that is 2D circulantly shifted by ${{r}}[m]$ rows and ${{c}}[m]$ columns of the base matrix $\mathbf{P}$.  To explain 2D-CCS, we consider a case where the BS applies all the $N^2$ possible distinct 2D circulant shifts of $\mathbf{P}$ to its antenna array, i.e., $M=N^2$. In this case, the measurement matrix $\mathbf{G}$ acquired by the receiver is the 2D-circular correlation between $\mathbf{H}$ and $\mathbf{P}$.  We denote the 2D circular cross-correlation as
\begin{equation}
\label{eq:defn_G}
\mathbf{G}=\mathbf{H}\ostar \mathbf{P},
\end{equation}
and the channel measurements become $\{y[k]\}_{k =1}^{N^2}$, which is a vectorization of $\mathbf{G}$. The $(r,c)^{\mathrm{th}}$ entry of $\mathbf{G}$ can be expressed as 
\begin{equation}
G(r,c)=\sum_{k,\ell}H(k,\ell)  \overline{P}\left((k-r)\, \mathrm{mod} \,N,(\ell-c)\, \mathrm{mod}\, N \right).
\end{equation}
In such a \emph{full sampling} case, i.e., $M = N^2$, we apply 2D-DFT on the two sides in (\ref{eq:defn_G}). Denoting the 2D-DFT of the measurement matrix $\rG$ as $\mathbf{F}$, we have 
\begin{align}\label{equ:DFT2Dcirc}
\underbrace{\rU_N^*\mathbf{G}\rU_N^*}_{\mathbf{F}}      = \rU_N^* (\rH\ostar{\rP})\rU_N^*  = \underbrace{\rU_N^*\rH\rU_N^*}_{\rX} \odot \underbrace{{N}\rU_N^*\rP_{\mathrm{FC}}\rU_N^*}_{\rZ}, 
\end{align}
where $\mathbf{X}$ is the beamspace defined in (\ref{equ:beamspace2D}) and $\mathbf{Z}$ is the 2D-DFT of the base matrix.

 Under the assumption that $\rZ$ is non-zero at all locations, the BS can recover the full beamspace channel by simply calculating $\mathbf{X}(i, j) = \mathbf{F}(i, j)/\mathbf{Z}(i, j), ~i, j \in [N]$. Acquiring all the $N^2$ entries of $\mathbf{G}$, however, results in a training overhead that is comparable to exhaustive search. In 2D-CCS, the BS applies fewer circulant shifts of $\rP$ compared to $N^2$ to reduce the overhead, i.e., $M\ll N^2$. We use $\Omega$ to denote an ordered set of $M$ circulant shifts used by the transmitter, i.e., $\Omega\{m\} = (r[m], c[m]), ~m \in \{1, 2, \cdots, M\}$. For example, $\Omega=\{(0,1), (1,2)\}$ is one possible set for $M=3$. In this case, the BS applies phase shift matrix is 2D circulantly shifted by $0$ rows and $1$ columns in the first training slot, and $1$ rows and $2$ columns in the second training slot. The receiver will correspondingly acquires $G(0,1),G(1,2)$. Hence, the channel measurements $\{y[m]\}_{m = 1}^{M}$ is simply a subsampling of the measurement matrix $\rG$ in (\ref{eq:defn_G}) at given locations in $\Omega$. 
 
  There are three key problems in designing 2D-CCS-based channel recovery: 1) how to design the base matrix $\rP$, 2) how to recover the channel with a subsampled version of $\mathbf{G}$, and 3) how to design the 2D circulant shift sequence $\{r[m]\}_{m = 1}^M$ and $\{c[m]\}_{m = 1}^M$. In this paper, we mainly investigate the first two problems and leave the 2D circulant shift optimization to future work. We assume that the 2D circulant shift sequences are selected randomly for tractability of analysis. To better explain the concepts of 2D-CCS, we define $\mathcal{P}_{\Omega}(\mathbf{G})$ as the vector containing the entries of $\mathbf{G}$ at the locations in $\Omega$. The compressed channel measurement vector in 2D-CCS is then
\begin{equation}
\label{eq:Subsamp_G}
\mathbf{y}=\mathcal{P}_{\Omega}(\mathbf{G})+\mathbf{v}.
\end{equation}
The measurements are called convolutional channel measurements as the subsampled cross-correlation operation in \eqref{eq:Subsamp_G} can be realized using subsampled convolution. For a well designed base matrix $\mathbf{P}$ in 2D-CCS, optimization algorithms can estimate the best beamspace direction from the compressed channel measurements in \eqref{eq:Subsamp_G} even when $M \ll N^2$ \cite{falp}.  
\par We define the beamspace $\mathbf{X}$ as the 2D-DFT of the channel $\mathbf{H}$ as in (\ref{equ:beamspace2D}) and $\mathbf{Z}[m]$ as the 2D-DFT of the phase shift matrix $\mathbf{P}[m]$ in (\ref{equ:DFT2Dcirc}). Given an ensemble of beamspace of different vehicular channel realizations, we define the beamspace prior of the $m$-th beam direction as the probability that the beam direction is the best beam within the 2D-DFT codebook, i.e., $|\rX[m]|\geq |\rX[j]|, \forall j \in \{1,2, \cdots, N^2\}$. The beamspace prior of our simulation setup in Fig. \ref{fig:simulationsetup} is shown in Fig. \ref{fig:beamspace_prior}.
 
To explain the efficiency of 2D-CCS, we consider an example where $\mathbf{P}[1]$ is set to a given base matrix, i.e., $\mathbf{P}[1]=\mathbf{P}$. The first channel measurement is then $\langle \mathbf{H}, \mathbf{P} \rangle$. Due to the unitary nature of the 2D-DFT, the inner product between $\mathbf{H}$ and $ \mathbf{P}$ can also be expressed as the inner product of their 2D-DFTs, i.e., $\langle \mathbf{X}, \mathbf{Z}[1] \rangle$. Thanks to the sparsity nature of mmWave channels, the information required for beam alignment is encoded in fewer coefficients of $\mathbf{X}$, i.e., the entries of $\mathbf{X}$ along the two strips. To efficiently extract this information from $\langle \mathbf{X}, \mathbf{Z}[1] \rangle$ in a way that is robust to noise, $\mathbf{Z}[1]$ must have a large amplitude at the locations along the two strips. A base matrix $\mathbf{P}$ whose 2D-DFT satisfies such a property can be used to efficiently acquire features or measurements of the sparse channel. Furthermore, an interesting property of 2D-CCS is that the matrices used to obtain channel projections have the same 2D-DFT magnitude as the base matrix, i.e., $|\mathbf{Z}[m]|=|\mathbf{Z}[1]| \, \forall m$. The observation follows from the fact that 2D-circulant shifts over a matrix do not change the magnitude of its 2D-DFT \cite{imageprocess}. Using this property, it can be observed that a careful design of the base matrix ensures that all the matrices $\{\mathbf{Z}[m]\}^M_{m=1}$ have a large amplitude at the desired locations and guarantees efficient measurements of the channels. Designing a base matrix that is best suited to the channel prior, however, can be challenging when $M\ll N^2$. In Section \ref{sec:DL_main}, we discuss how deep learning can be used to optimize the base matrix for efficient 2D-CCS.

 \subsection{2D-CCS optimization using deep learning} \label{sec:DL_main}
As discussed in Section \ref{sec:2dccs}, channel measurements acquired in 2D-CCS are essentially a subsampling over the 2D circular correlation between the channel and the base matrix. Such a convolutional sensing structure in 2D-CCS makes it a desirable candidate to be optimized through deep learning. In this section, we discuss how the deep neural networks can be constructed to emulate different components of 2D-CCS and used to optimize 2D-CCS-based beam alignment. 

\par
\begin{figure*}[h!]
\centering
\includegraphics[trim = 0cm 9cm 0cm 0cm, clip = true, width=1.0\textwidth]{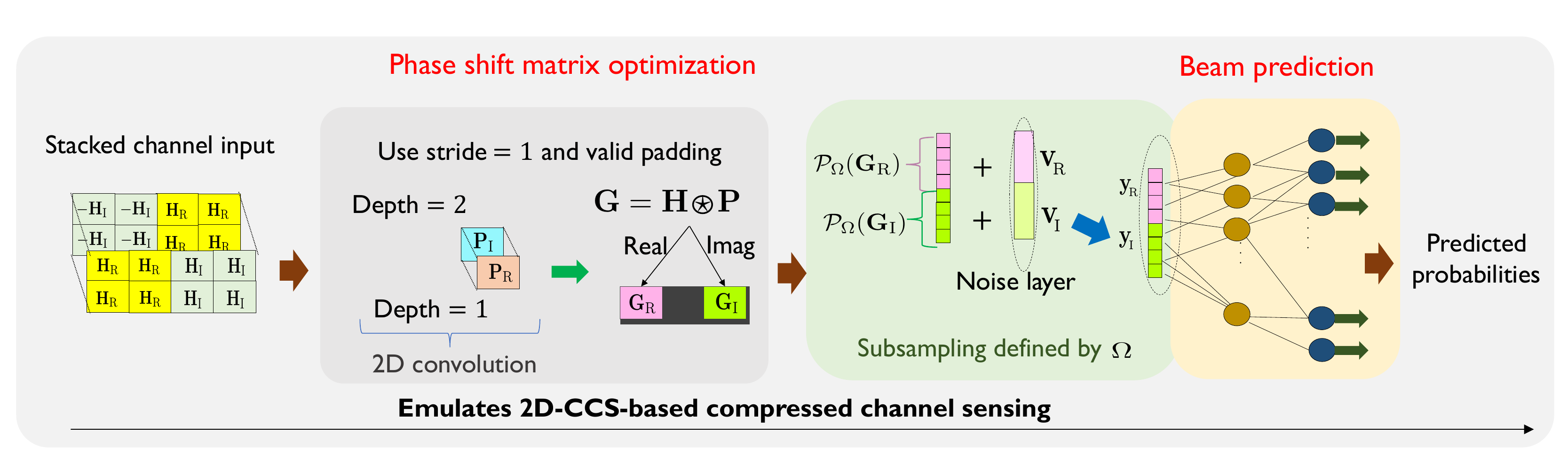}
\caption{ Channel measurements in 2D-CCS are realized using convolutional filters $\mathbf{P}_{\mathrm{R}}$ and $\mathbf{P}_{\mathrm{I}}$. Using end-to-end learning, the base matrix in 2D-CCS, i.e.,  $\mathbf{P}=\mathbf{P}_{\mathrm{R}}+\mathsf{j} \mathbf{P}_{\mathrm{I}}$, is optimized for compressively measuring the channel and maximize the probability of beam alignment. The fully connected layers are optimized to predict the optimal beam direction based on the compressive channel measurements.}\
\normalsize
  \label{fig:2dccs_illus}
\end{figure*}

 We explain the proposed deep neural network architecture using Fig. \ref{fig:2dccs_illus}. The first part of our deep neural network contains a 2D complex convolutional layer to emulate the full 2D-CCS subsampling in (\ref{eq:defn_G}). The weights of the convolutional filter model the real and imaginary parts of the base matrix $\mathbf{P}$ in 2D-CCS. The next is a subsampling layer that subsamples the full sampling measurement $\rG$ in (\ref{eq:defn_G}), and obtains 2D-CCS-based compressed channel measurements given the subsampling set $\{r[m]\}_{m = 1}^M$ and $\{c[m]\}_{m = 1}^M$. The last component of the network is a cascade of fully-connected layers that predicts the best beam from the compressed channel measurements obtained in (\ref{eq:Subsamp_G}). Using a complex CNN, our network targets at optimizing parameters of two components in 2D-CCS. First, the optimized weight of the 2D convolutional filter is the base matrix $\mathbf{P}$ that is implemented at the transmitter phased array; 2) the weights of the fully-connected layers are optimized to predict the optimal beam index based on the compressive channel measurements acquired by 2D circulant shifts of $\rP$.  In this section, we describe the key components of our network and the training procedure. Our implementation is available online on GitHub \cite{Code_deep_learning}.
\subsubsection{2D-CCS emulation using complex CNN}\label{sec:2dccs_implem}
\par
\begin{figure}[ht]
\centering
\includegraphics[width = 5.0in]{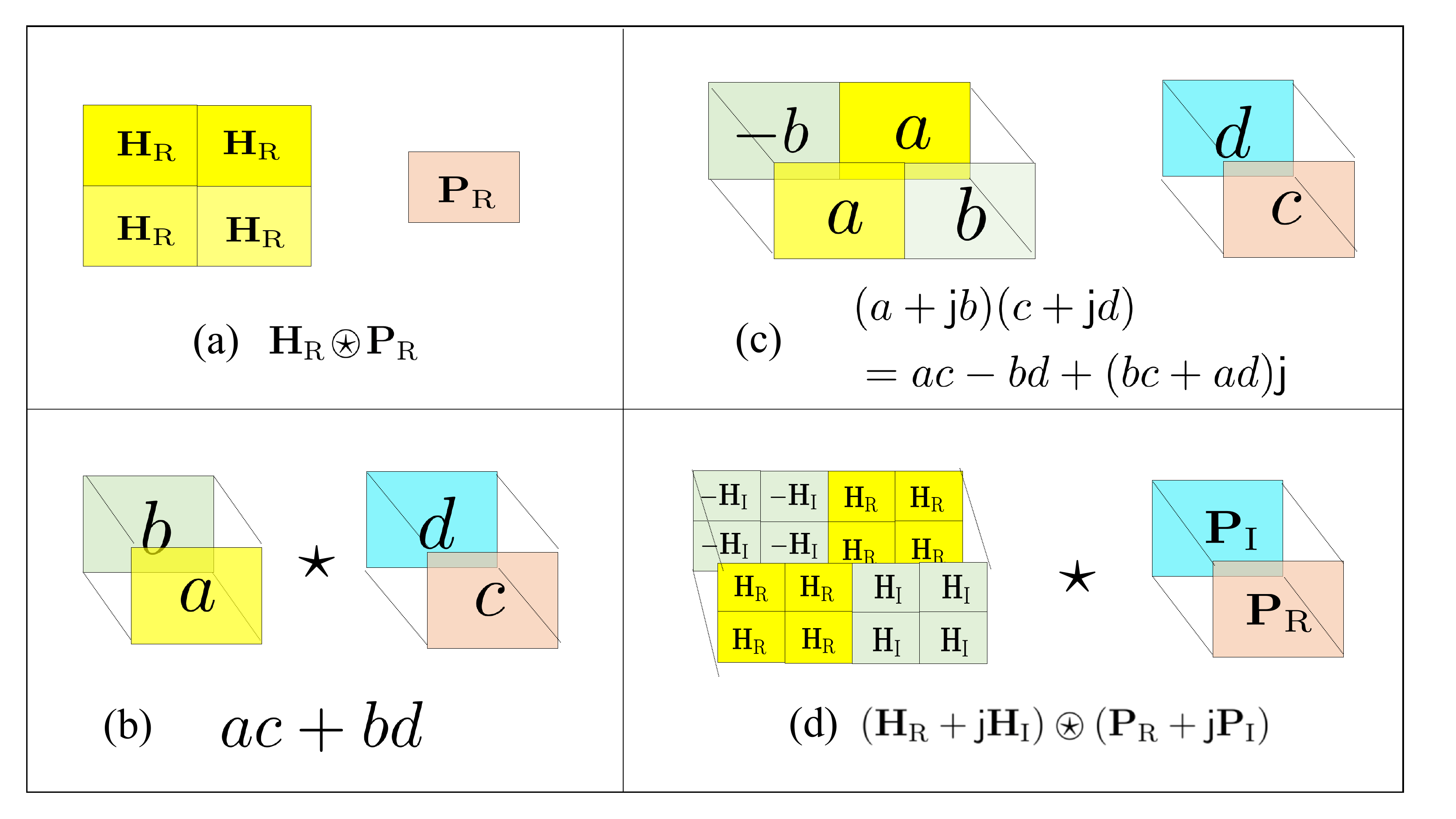}
\caption{An illustration of how to implement 2D circular convolution using CNN. In Fig. \ref{fig:convillus}a, we show an implementation of circular convolution of two real matrices $\rH_\mathrm R$ and $\rP_\mathrm R$ by repeating $\rH_\mathrm R$. Fig.  \ref{fig:convillus}b demonstrates an example of a 2-channel CNN. The result sums up the product of $ac$ and $bd$. Fig.  \ref{fig:convillus}c shows an example that a 2-channel CNN can achieve the product of two complex variables $a + \mathsf{j}b$ and $c + \mathsf{j}d$, where the left part of the correlation in Fig. \ref{fig:convillus}c is the real component and the right is imaginary. Lastly, Fig.  \ref{fig:convillus}d combines the idea in Fig.  \ref{fig:convillus}b and Fig.  \ref{fig:convillus}c and demonstrates how to implement 2D circular convolution between two complex matrices.  }\label{fig:convillus}
\end{figure}

 In this section, we explain how to emulate 2D-CCS and implement its different components using a CNN. To obtain the complex 2D circular correlation in \eqref{eq:defn_G}, we need to address two issues: 1) implementation of 2D circular correlation, and 2) realizing correlation of complex matrices using a real-valued CNN. We denote $\mathbf{H}_{\mathrm{R}}$ as the real part of the channel, $\mathbf{P}_{\mathrm{R}}$ as the real part of the  matrix, and accordingly the imaginary part $\rH_\mathrm R$ and $\rP_\mathrm I$. To realize a circulant structure in the correlation, we define a matrix $\mathbf{A}_{\mathrm{R,pad}}=[\mathbf{A}_{\mathrm{R}},\mathbf{A}_{\mathrm{R}};\mathbf{A}_{\mathrm{R}},\mathbf{A}_{\mathrm{R}}]$. Note that $\mathbf{H}_{\mathrm{R,pad}}$ is a $2N \times 2N$ matrix. As illustrated in Fig. \ref{fig:convillus}a, applying the convolutional filter $\mathbf{P}_{\mathrm{R}}$ over $\mathbf{H}_{\mathrm{R,pad}}$, in a valid-padding mode with a stride of $1$, results in an $(N+1)\times (N+1)$ matrix. The circular cross-correlation $\mathbf{H}_{\mathrm{R}} \ostar \mathbf{P}_{\mathrm{R}}$ is obtained by simply deleting the last row and the last column of the convolved output. In Fig. \ref{fig:convillus}b and Fig. \ref{fig:convillus}c, we explain the implementation of complex correlation using CNN. For 2-channel convolutional networks as shown in Fig. \ref{fig:convillus}b, the output of the convolutional layer adds up the correlation of the matrices on the first channel and the second channel respectively. Leveraging the computation property of multichannel convolution in Fig. \ref{fig:convillus}b, the complex correlation can be calculated by expanding the input matrix as Fig. \ref{fig:convillus}c, where proper construction of the real and imaginary parts of complex variables can achieve a product of two complex values. Finally, combining the construction rules in Fig. \ref{fig:convillus}a and  Fig. \ref{fig:convillus}c, we apply the Fig. \ref{fig:convillus}d to obtain the 2D circular correlation $\rH \ostar \rP$, between the complex channel $\rH$ and the complex base matrix $\rP$. Stacking the channel as in Fig. \ref{fig:convillus}, we obtain the inputs for the deep learning model that outputs the full channel measurement in $\rG = \rH\ostar \rP$ after the convolutional layer. 

 Channel stacking in Fig. \ref{fig:convillus} constructs a $2N\times 4N \times 2$ real-valued tensor from each complex channel $\mathbf{H}$. An $N\times N \times 2$ convolutional filter with $\mathbf{P}_{\mathrm{R}}$ and $\mathbf{P}_{\mathrm{I}}$ as the filter weight  is applied over the input channel tensor. The filtering is performed in a valid-padding mode with a stride of $1$ to result in an $(N+1)\times (3N+1)$ matrix. The element-wise correlation between the $2N\times 4N \times 2$-dimensional real valued tensor and the $N\times N\times 2$ filter provides a the desired results $\rG_\mathrm R$ and $\rG_\mathrm I$, which are the real and imaginary components of the complex matrix correlations $\rH \ostar \rP$. Finally, $\mathbf{G}_{\mathrm{R}}$ and $\mathbf{G}_{\mathrm{I}}$ are $N \times N$ matrices which begin at the $(0,0)$ and $(2N,0)$ coordinates of the $(N+1)\times (3N+1)$ matrix.
\par The channel measurements in \eqref{eq:Subsamp_G} can be split into real and imaginary components as $\mathbf{y}_{\mathrm{R}}=\mathcal{P}_{\Omega}(\mathbf{G}_{\mathrm{R}})+\mathbf{v}_{\mathrm{R}}$ and $\mathbf{y}_{\mathrm{I}}=\mathcal{P}_{\Omega}(\mathbf{G}_{\mathrm{I}})+\mathbf{v}_{\mathrm{I}}$. We keep only the real and imaginary parts $\rG_\mathrm R$ and $\rG_\mathrm I$ of the convolutional layer output by ignoring the irrelevant nodes. Then, we implement  a subsampling over the full measurement $\rG$ by using an identical dropout technique for both $\mathbf{G}_{\mathrm{R}}$ and $\mathbf{G}_{\mathrm{I}}$. Our dropout procedure discards $N^2-M$ entries of $\mathbf{G}_{\mathrm{R}}$ and $\mathbf{G}_{\mathrm{I}}$ at the same locations in the set $\Omega$ which are chosen at random. The subsampled features, i.e., $\mathcal{P}_{\Omega}(\mathbf{G}_{\mathrm{R}})$ and $\mathcal{P}_{\Omega}(\mathbf{G}_{\mathrm{I}})$, are perturbed by AWGN to emulate the compressed sensing model in \eqref{eq:Subsamp_G}. The subsampled channel feature vector of dimension $2M \times 1$ is fed into a beam prediction network that is a cascade of fully-connected layers. ReLU activation is used at all the fully-connected layers and a softmax is used at the output layer for beam index classification. In this paper, beam alignment is performed using a 2D-DFT codebook with a total of $N^2$ elements. Each codebook element is modeled by a class and the output of the neural network is an $N^2 \times 1$ vector that contains the predicted class probabilities. 
The proposed deep learning model mimics a complete 2D-CCS-based beam alignment framework for mmWave vehicular communication. After the offline learning, we can configure the base matrix $\rP$ at the transmitter phased array using the convolutional filter weight. The receiver acquires a sequence of channel measurements using the 2D circulant shifts of the phase shift matrix based on $\Omega$. Finally, the BS can directly predict the optimal transmit beam based on the compressive measurement feedback based on the optimized fully-connected layer weights. 

\par The network is trained using the restructured channels, i.e., a collection of $2N \times 4N \times 2$ tensors defined in Section \ref{sec:2dccs_implem}, and the best beam class indices corresponding to the channels. We denote the number of samples in the dataset as $K$. We define the input training dataset as $\{{\boldsymbol{x}}_k\}_{k = 1}^K$ and the output label in the training dataset as $\{{\boldsymbol{y}}_k\}_{k = 1}^K$. We define $f: \mathbb{R}^k \to n$ as the parameters learned in the neural networks. The predicted output is therefore  $\{{\boldsymbol{{\hat{y}}}}_k: ~{\boldsymbol{{\hat{y}}}}_k = f({\boldsymbol{x}}_k)\}_{k = 1}^K$. Defining $\ell(\hat{\boldsymbol{y}}, \boldsymbol{y})$ as the negative log likelihood of $x$, the loss functoin $J(\cdot)$ is defined as 
\begin{align}\label{equ:loss}
J(\{{\boldsymbol{x}}_k\}_{k = 1}^K, \{{\boldsymbol{y}}_k\}_{k = 1}^K) = \frac{1}{K}\sum_{k = 1}^K \ell (\boldsymbol{{y}}_k, f(\boldsymbol{x}_k)).
\end{align} The deep neural network is trained by minimizing the cross entropy between the predicted class probabilities $\{{\boldsymbol{{\hat{y}}}}_k\}_{k = 1}^K$ and the one-hot encoded labels $\{{\boldsymbol{{{y}}}}_k\}_{k = 1}^K$ corresponding to the best beam index. 
\subsubsection{Optimization with low-resolution phase shifters}\label{sec:2dccs_quant_implem}
In Section \ref{sec:2dccs_implem}, we introduced the deep learning-based framework to optimize 2D-CCS base matrix, and the fully-connected layers whose weights are used to predict the optimal transmit beam. Such a framework, however, does not impose constraints over the phase shift matrix to be optimized, i.e., the phase shift matrix $\rP\in\mathbb{C}^{N\times N}$. In reality, the phase shift matrix is subject to several constraints due to the limitations of hardware. First, the phase shift matrices follow power constraints. It is feasible to impose unit-norm constraint over the phase shift matrix, i.e., $\Vert\rP\Vert_\mathrm{F} = N$. For example, \texttt{tf.keras.constraints.UnitNorm} function in \texttt{Tensorflow}
 constrains the weights to be unit-norm during training. Such unit-norm constraint, however, is not enough. The phase shift matrix needs to be unit-norm per phase shifter, i.e., $|\rP| = \mathbf{1}$ during the training since the real and imaginary parts of $\rP$ are split in the 2-channel convolutional filters for optimization.
Second, phase shifters are subject to low-resolution constraints, i.e., $\rP\in \mathbb{Q}_q^{N\times N}$. To keep the low-resolution constraints of the phase shifters during the training, an intuitive solution is to directly add an extra layer which applies quantization over the filter of the convolutional layer. Directly applying optimization with quantization in training, however, is infeasible since the quantization function is not differentiable and will hinder the weight update calculation in backpropagation. 

In this paper, we propose a projected gradient descent-based solution to include hardware constraints in the convolutional filter optimization during the training \cite{boyd2004convex, bubeck2014convex, grubb2010boosted}. With a convex loss function, stochastic gradient descent, an iterative method, is leveraged to optimize the parameters with suitable smoothness properties in deep neural networks \cite{le2011optimization, hecht1992theory, schmidhuber2015deep}. To make the phased shift matrix hardware-compatible, the convolutional filters are subject to the quantization constraint given the resolution of the phase shifters. Projected gradient descent is a natural solution to optimize a convex problem given a feasible set. It is a simple variation of stochastic gradient descent by adding an extra procedure of projecting the variable onto the feasible set for each update. In our paper, the projection function is simply a quantization function of the phase shifter's phase, which we denote as $\mathcal{Q}_q(\cdot)$ given the resolution as $q$. Given a starting base matrix at step $k$ as $\rP^{(k)}\in\mathbb{Q}$, step size as $\alpha$, projected gradient descent works as follows until certain stopping criteria are met, 
\begin{align}
\rP^{(k + 1)} =\mathcal{Q}_q\left({\rP^{(k)}) - \alpha\nabla f(\rP^{(k)})}\right),
\end{align}
where $\nabla f(\rP^{(k)})$ denotes the gradient of the convolutional filter calculated in backpropagation. In this paper, we adopt a uniform quantization function. Denoting $\Delta = 2\pi/2^q$ and $\angle(\cdot)$ as the phase of the complex variable, we define the quantization function $\mathcal{Q}(\cdot)$ over the phase shift matrix as \begin{align}\label{equ:quant}
\mathcal{Q}_q(\rP): ~P(r, c) = e^{\mathsf{j}\Delta\floor{\frac{\angle \left(P(r, c)\right)}{\Delta}} }, ~\forall r, c \in\{1, 2, \cdots, N\}. 
\end{align}

Projected gradient descent is a simple approach to impose constraints on the network parameters. It pulls the updated phase shift matrix to the feasible set $\mathbb{Q}_q$ after each weight update and guarantees the convolutional filter is hardware-compatible during the training. For the proposed CNN model, such projected gradient descent-based quantization only impacts the calculation of the forward propagation loss. The propagation loss further changes the gradients applies convolutional filter weight update subject to constraint. In this paper, we apply quantization to the convolutional filter weights every $N_\mathrm c > 1$ mini-batches. The reason to quantize weights every few mini-batches is that backpropagation could be stuck at local minimums. For example, the norm of the gradient update could be very small, i.e.,  $\Vert\alpha \nabla f(x^{(k)})\Vert_\mathrm F\ll \Vert\rP^{(k)}\Vert_\mathrm F$, which leads to $\rP^{(k + 1)}
\approx \rP^{(k)}$. During the training, we keep the Frobenius norm of the phase shift matrix to be constant, and focus on the optimization of the phase shift matrix's phase only, which is the target of the work. We impose identical per-entry norm of $\rP$, i.e., $|\rP| = {\omega}\mathbf{1}$, where $\omega$ is the Frobenius norm of the phase shift matrix $\rP_0$ that is initialized randomly $\omega = \Vert\rP_0\Vert_\mathrm F$. After the offline learning, the optimized phase shift matrix with low-resolution constraint is normalized to satisfy the unit-norm power constraint $\rP = N\rP/\omega$. The proposed approach is summarized as Algorithm \ref{alg:algorithm1}.  
\begin{algorithm}[ht]
  \caption{Training a CNN with low-resolution convolutional filters}\label{alg:algorithm1}
  \begin{algorithmic}[1]
    \Require
      \State A mini-batch of inputs $N_\mathrm b$, the intervals $N_\mathrm c$ for projected gradient descent-based weight update, phase shifter resolution $q$ and the quantization function $\mathcal{Q}_q(\cdot)$ defined in (\ref{equ:quant}), the training inputs $\boldsymbol{x}$ and the training outputs $\boldsymbol{y}$, the loss function defined as $J(\hat{\boldsymbol{y}}; \boldsymbol{y})$ in (\ref{equ:loss}), where $\hat{\boldsymbol{y}}$ is the predicted outputs obtained by $f(\boldsymbol{x})$.\\
    \textit{Initialization}: Initialize random weight parameters $\rW_\mathrm{conv}$ and $\rW_\mathrm{fc}$, initialize Frobenius norm of the weight as 
    $\omega_\mathrm{conv} = \Vert\rW_\mathrm{conv}\Vert_\mathrm F$. 
          \For {$t = 0: T_\mathrm b - 1 $}
          \If {$t$ mod $T_c == 0$}
          \State $\rW = \rW_\mathrm{conv}[:, :, 0] + \mathsf{j}\rW_\mathrm{conv}[:, :, 1]$ \Comment {Construct the phase shift matrix.}
         \State $\rW_\mathrm{conv} = \mathcal{Q}_q(\rW)$ \Comment{Quantize phase shift matrix $\rW$. }
         \State $\rW_\mathrm{conv} =  \omega_\mathrm{conv} \rW_\mathrm{conv}$. \Comment{Keep the norm of the phase shift matrix constant. }
                  \EndIf
 \State Calculate forward propagation loss based on loss function $C = J(f(\boldsymbol{x}); \boldsymbol{y})$. 
         \State Calculate gradients in backpropagation $g_{\mathrm{conv}} = \frac{\partial C}{\partial \rW_\mathrm{conv}}$ and $g_{\mathrm{fc}} = \frac{\partial C}{\partial \rW_\mathrm{fc}}$
          \State Updates the convolutional filter weight $\rW_\mathrm{conv}$, and fully connected layers weights $\rW_\mathrm{fc}$ based on the gradients $g_{\mathrm{conv}}$ and $g_{\mathrm{fc}}$. 
                       \EndFor
                       \State $\rW_\mathrm{conv} = \rW_\mathrm{conv}/\Vert\rW_\mathrm{conv} \Vert_\mathrm F$. \Comment{Normalize the phase shift matrix with unit-norm constraint.}
  \end{algorithmic}
\end{algorithm}

\subsubsection{Feature normalization and model retraining}\label{sec:scaling}
\begin{figure}
	\centering
	\includegraphics[trim = 0cm 0.5cm 0cm 0.5cm,clip=true, width = 4.5in]{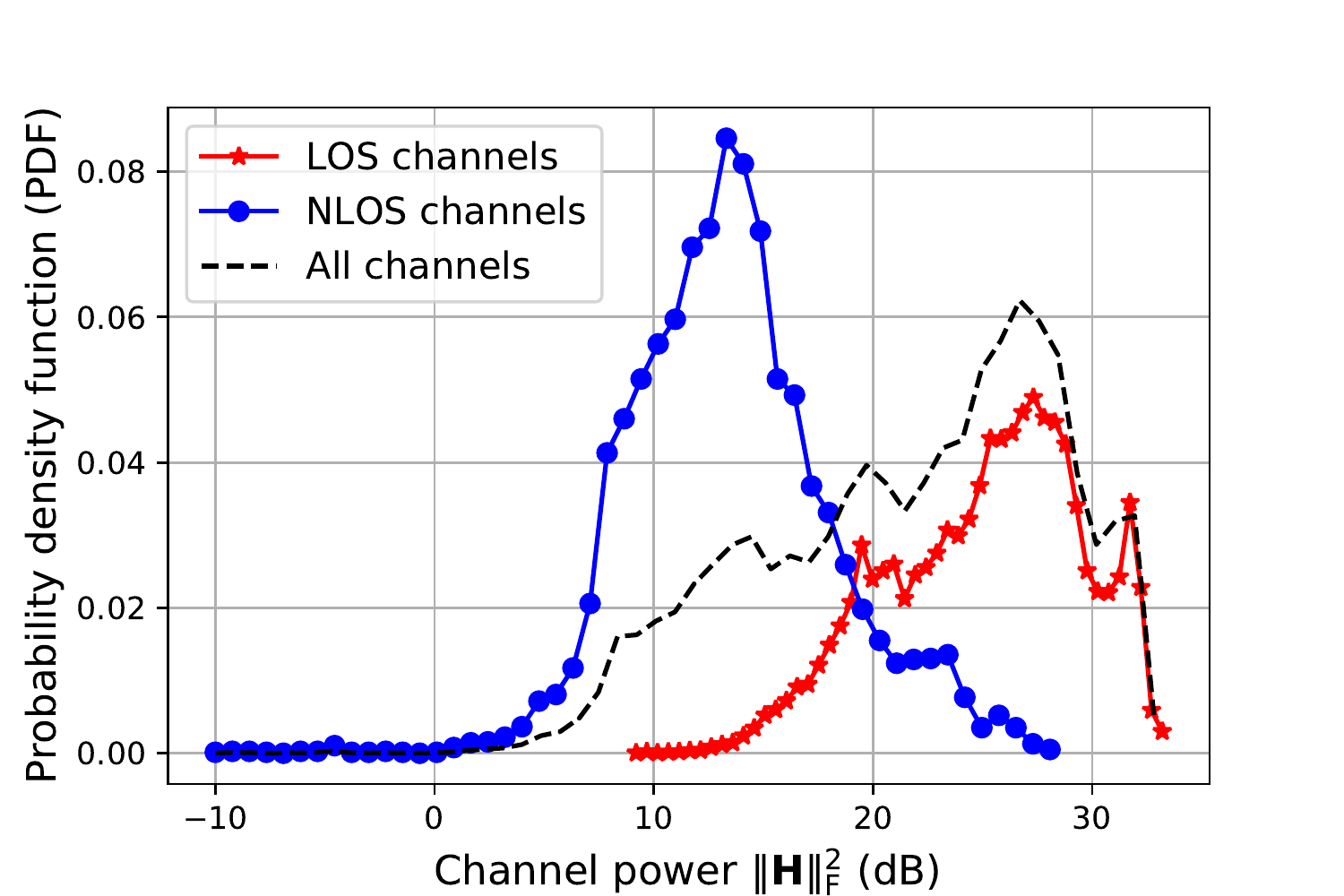}
	\caption{A comparison of the PDF of powers of the channels $\Vert\rH\Vert_\mathrm F^2$.  The black dashed line represents the PDF of all data samples (including LOS/NLOS). The red curve with stars represents the PDF of the LOS channels whose direct LOS path is free of blockage channels. The blue curve with solid circle illustrates the PDF of the NLOS channels, where the LOS path is blocked by vehicles such as the truck.  }\label{fig:powerdistribution}
\end{figure}

We calculate the channel model based on (\ref{eq:channel_expansion}) given the ray tracing simulation outputs. The channels of different vehicles might have different pathloss, based on their relative distance to the BS, the physical location, and if the LOS path is blocked, etc. In the dataset, we have an ensemble of vehicular channels with vehicles randomly distributed in a street of length $L  = 100$ m. In Fig. \ref{fig:powerdistribution}, we show the probability density function (PDF) of the channel power $\Vert\rH\Vert^2_\mathrm{F}$ of the entire dataset, that contains LOS channels only, and NLOS channels only. The whole dataset consists of a total of $26$K channels, among which $19$K are LOS and the rest $7$K channels are NLOS. The average power for LOS channels is $25.4$ dB; the average power for NLOS channels is $14.0$ dB; while the average power for all channel samples is $22.1$ dB. In Fig. \ref{fig:powerdistribution}, the red curve with stars represents the LOS channels free of blockage and the blue curve with circles represents the NLOS channels where the direct LOS path is blocked by trucks. The dashed line illustrates the PDF of channel power of the ensemble of all channels, including the LOS and NLOS channels. We can observe that there exists a big difference between the power of LOS and NLOS norms. 

Feature scaling is a method used to normalize the range of features, which is also referred to as data normalization \cite{juszczak2002feature}. Properly normalized features guarantee that independent features contribute proportionately to the outputs \cite{ioffe2015batch}, \cite{aksoy2001feature}. Furthermore, the feature normalization improves the speed of convergence based on gradient descent. As the power of the channels is significantly different, as can be observed from Fig. \ref{fig:powerdistribution}. There are several challenges for the deep neural networks to train using the channel inputs with a large variance in the Frobenius norm. First, different training channel inputs with a wide feature range lead to slower gradient descent optimization. Second, the channels with large variance makes it challenging for the phase shift matrix optimization. The proposed deep learning model targets at optimizing a phase shift matrix that maximizes the predicted beam alignment probability. Therefore, the 2D-DFT of the phase shift matrix $\rZ$ will have a significantly larger magnitude at those beam directions that have a higher probability to be optimal. The strongest beam direction of those NLOS channels will not be allocated enough power for measurement since it is less likely that they are going to be recovered and contribute to reducing the training loss. As a result, the optimization will lead to a poorly-designed mask with power mainly focused on the \emph{popular} beam directions corresponding to the channels with high power. Such a mask will fail at recover the strongest beam direction associated with the NLOS channels. 

In this paper, we preprocess the data and adjust the training procedures to minimize the effects of large channel power variance on the model prediction accuracy as follows. We propose a \textbf{two-stage} training framework to optimize the phase shift matrix, i.e., the mask in the first stage, and the fully-connected layers for beam prediction in the second stage. 
\begin{itemize}
	\item \textit{\textbf{Stage 1:} base matrix optimization}\\
In the first stage, to guarantee a mask that has a reasonable magnitude in the \emph{less popular} beam directions, we first normalize \emph{all} training channels with unit-norm power constraint, i.e., $\rH \to \rH/\Vert\rH\Vert_\mathrm F$. This removes the effect of large variance in the training input. Note that a well-designed mask is one that covers all \emph{important} beam directions. Therefore, we neglect the effect of noise to make sure that all beam directions can be successfully recovered for mask optimization in the training to successfully recover the the optimal beam directions. With training channel inputs of unit power constraint and the noise-free setup, the model is able to properly optimize the phase shift matrix. The 2D-DFT representation, the mask $\rZ$ covers all different beam directions with reasonable magnitudes, regardless of the channels with large power variance and likely bad SNRs. The \textbf{Stage 1} \emph{only} optimizes the phase shift matrix. 
\item \textit{\textbf{Stage 2:} beam alignment optimization}
\\
Removal of the large variance effects of the channel data improves the chance of recovering all important beam directions. It should be noted, however, that the real-world channel realizations do not have such desired properties.  Instead, the channels vary a lot across samples and there be could be a lot of outages due to low channel SNRs. Such a difference will not have an effect on the phase shift matrix optimization since the phase shift matrix is primarily used to acquire channel in an efficient way, but it will lead to catastrophic evaluation performance when it comes to the optimization of the beam alignment optimization. Hence, we propose to retrain the fully-connected layers after fixing the phase shift matrix optimized in \textit{\textbf{Stage 1}}. We normalize both the training and testing channels given expected unit-norm constraint $\mathbb{E}[\Vert\rH\Vert_\mathrm F^2] = N$. 
\end{itemize}
With retraining, we are able to obtain an optimized model with well-designed mask that is able to recover important beam directions despite the large channel power variance (in \textbf{Stage 1}), and fully-connected layers that are suited to the realistic channel realizations (in \textbf{Stage 2}).
\section{2D-CCS optimization with wideband channels}\label{sec:wideband}
In Section \ref{sec:DL_main}, we explained how a properly designed CNN can be applied to mimic 2D-CCS predict beam alignment using narrowband channels. In this section, we answer the question of how the narrowband channel-based framework can be extended to frequency-selective channels at mmWave. When the wideband channels are available, is it possible to design a better phase shift matrix and achieve higher beam alignment accuracy? 
\subsection{Input structure of wideband channels}
To explain the extension of the CNN model to wideband channels, we first introduce the wideband channel model. We define the pulse-shaping filter as $g(\cdot)$, the sampling period as $T$. Based on the geometric channel model, the wideband channel of $L_\mathrm c$ taps can be written as 
\begin{align}
\label{eq:channel_expansion}
\mathbf{H}[n]=\sqrt{N_\mathrm t N_\mathrm r}\sum_{\ell = 0}^{L_\mathrm{p} - 1}g(nT - \tau_\ell) \alpha_\ell e^{\mathsf{j}\beta_\ell}\mathbf{a}_N(\cos\theta_\ell)\mathbf{a}_N{(\sin\theta_\ell\cos\phi_\ell)}^T, ~0\leq n \leq L_\mathrm c - 1
\end{align}
We calculate the discrete Fourier transform of the multi-tap channels, along the different taps to obtain the subcarriers $\mathbf{\mathsf{H}}[0], \mathbf{\mathsf{H}}[1],$ $\cdots, \bold{\mathsf{H}}[L_\mathrm c - 1]\in\mathbb{C}^{N\times N}$. In simulation setup, we set $L_\mathrm c = 128$. To avoid training inputs redundancy, we subsample every $L_\mathrm{sub} = 8$ subcarriers to reduce correlation of adjacent subcarriers. In the downlink of LTE, there is an unused DC subcarrier, i.e., $\mathsf{H}[0]$ that coincides with the carrier center frequency. The DC subcarrier is zeroed out in downlink transmission because it is subject to ultra high interference due to, for example, local-oscillator leakage. Hence, we subsample the subcarriers from the first subcarrier $\mathsf{H}[1]$, which gives a set of subcarriers as $\mathcal{{H}} = \{\mathsf{H}[sL_\mathrm{sub} + 1], s\in\mathcal{S}\}$, where $\mathcal{S} = \{0, 1, \cdots, 15\}$. To incorporate the wideband channels within 2D-CCS, we propose to concatenate the stacked wideband channels for training input. For each subcarrier, we prepare the input using $\mathsf{H}[sL_\mathrm{sub} + 1]\in\mathbb{C}^{N\times N}$ based on Fig. \ref{fig:convillus} to $\bar{\mathsf{H}}[s]\in\mathbb{C}^{2 \times 2N\times 4N},~ s \in \{0, 1, \cdots, 15\}$. Then the inputs for each channel tap are concatenated along the different channel taps to form $\bar{\mathsf{H}} = \large[\bar{\mathsf{H}}[0], \bar{\mathsf{H}}[1], \cdots, \bar{\mathsf{H}}[15]\large] \in \mathbb{C}^{2\times 2N \times 4NN_\mathrm{sc}}$, where we assume $N_\mathrm{sc} = |\mathcal{S}| = L_\mathrm c/L_\mathrm{sub} = 16$ in this paper.

\begin{figure}
\centering
\includegraphics[trim = 0cm 1cm 0cm 0cm, width =1.0\textwidth]{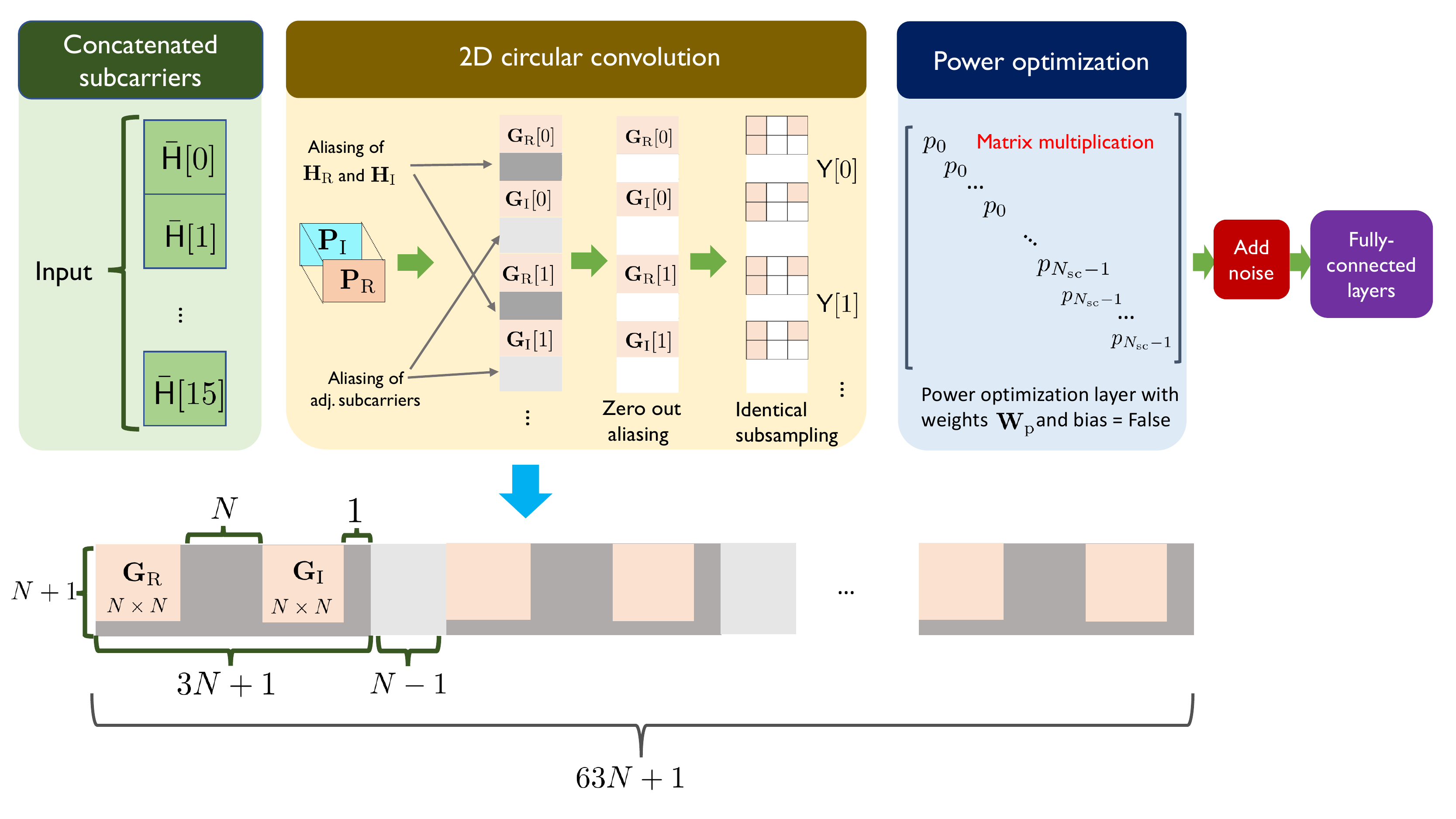}
\caption{An illustration of implementing 2D-CCS with multi-subcarriers. Each subcarrier $\mathcal{{H}} = \{\mathsf{H}[sL_\mathrm{sub} + 1]\}$ is stacked as Fig. \ref{fig:convillus}d to form $\bar{\mathsf{H}}[s], ~s\in\{0, 1, \cdots, N_\mathrm{sc} - 1\}$. The collection of $\bar{\mathsf{H}}[s], ~s\in\{0, 1, \cdots, N_\mathrm{sc}\}$ is then concatenated along the different subcarriers and fed into the convolutional layer whose filter has a weight of $\rP$. The output of the convolutional layer provides a concatenation of  2D circular correlation of individual subcarriers and the phase shift matrix. We zero out the irrelevant output units after the 2D circular correlation in the convolutional layer. Then, we apply optimization over the power allocated to different subcarriers using a linear layer, but enforcing the weight matrix to be diagonal and the bias to be false. The output is then added by the noise and fed into the fully-connected layer for beam alignment optimization. }\label{fig:wideband}
\end{figure}
An illustration of the proposed deep learning model in Fig. \ref{fig:2dccs_illus} to wideband channels is provided in Fig. \ref{fig:wideband}. First, given the stacked representation of $\bar{\mathsf{H}}[s],~s \in \{0, 1, \cdots, N_\mathrm{sc} - 1\}$, we concatenate the stacked representations along the different subcarriers. The concatenated subcarrier input is then fed to the 2D convolutional layer whose weight is $\rP$. It should be noted that the phase shift matrix to optimize is shared across different subcarriers, due to the fact that the analog beamforming matrix is the same for different subcarriers. The output of each subcarrier after the convolutional layer includes the real part, the \emph{aliasing} caused by the correlation between $\rP$ and the \emph{border} of $\rH_\mathrm R$ and $\rH_\mathrm I$ denoted by \emph{dark gray}, and the imaginary part. Between the outputs of subcarrier correlation, there is \emph{aliasing} caused by the 2D circular correlation of adjacent subcarriers, denoted by \emph{light gray}. We design a layer that zeroes out the aliasing and applies identical subsampling for the real and imaginary part of different subcarriers.
The subsampling trajectories are updated correspondingly to form the identical subsampling for different subcarriers, where we repeat the subsampling vector in the narrowband case along the channel tap axis to maintain the consistency of measurement subsampling for different subcarriers. In this case, the convolutional and subsampling layers output the compressive channel measurements acquired at different subcarriers and these measurements are fed in the fully-connected layers to predict the optimal beam index for beam alignment. 
\subsection{Power allocation of different channel taps}\label{sec:pow_alloc}
We assume the total transmit power is 1. The power allocated for the $n$-th subcarrier is $p_s^2,~\sum_{s \in\mathcal{S}} p_{k}^2= 1$. We have two baseline models for comparison. In the first scenario, the model allocates all power to the first subcarrier as input, i.e., $\mathsf{H}[1]$, which is equivalent to the narrowband case. In the second scenario, the model allocates the power equally to different subcarriers, which gives $p_s^2 = 1/N_\mathrm{sc}, ~\forall s \in \mathcal{S}$. It should be noted that we only consider amplitude scaling of the subcarrier in this paper, and hence $p_s,~\forall s\in\mathcal{S}$ is real.

We now propose deep learning architecture to optimize the power allocated to different subcarriers. Since the measurements at the receiver is distorted by the noise, the power allocation optimization to different subcarriers has to be applied before the noise adding layer in Fig. \ref{fig:wideband}.  To realize power allocation across different subcarriers, we add a \emph{linear} layer after the identical subsampling. First, we permute the axis of $\bar{\mathsf{H}}$ to $\bar{\mathsf{H}}\in \mathbb{R}^{2  \times 2N \times 64N}$. Hence, the output after the 2D convolutional layer becomes $\bar{\mathsf{Y}}_\mathrm{CNN}\in \mathbb{R}^{(N + 1) \times (63N + 1)}$. The sizes of the measurement block and aliasing are illustrated in Fig. \ref{fig:wideband}. After the convolutional layer and identical subsampling, we apply a linear layer with weight defined as $\rW_\mathrm P\in \mathbb{R}^{(63N + 1)\times (63N + 1)}$, with the bias defined as false. Therefore, the output after the linear layer is simply the multiplication between $\bar{\mathsf{Y}}_\mathrm{CNN}$ and $\rW_\mathrm{opt}$, which gives $\bar{\mathsf{Y}}_\mathrm{opt} =\bar{\mathsf{Y}}_\mathrm{CNN}\rW_\mathrm{opt}$, and $\bar{\mathsf{Y}}_\mathrm{opt}\in \mathbb{R}^{(N + 1) \times (63N + 1)}$. To apply identical scalars to the measurements acquired with the same subcarrier $\bar{\mathsf{H}}[s]$,  the optimization matrix $\rW_\mathrm{opt}$ should be a \emph{block} diagonal matrix, with the $s$-th block being an identity matrix multiplied by the scalar $p_\mathrm s$. We update the weight matrix based on the block diagonal constraint during the training and explain the algorithm in Algorithm \ref{alg:algorithm2}.\begin{algorithm}[ht]
	\caption{ Weight update to apply amplitude scalings for different subcarriers}\label{alg:algorithm2}
	\begin{algorithmic}[1]
		\Require\State The weight $\rW_\mathrm{p}$ of the optimization layer after weight update in backpropagation.
		\State $\rW = \mathrm{Diag}(\rW_\mathrm{p})$ \Comment{Diagonalize weight matrix.}
		\For{$s = 0$ to $N_\mathrm{sc} - 2$} 
		\State
		$w_s = \sum_{k = 4Ns}^{4Ns + 3N} |W_\mathrm{p}(k, k)|/(3N + 1)$
		\Comment{Calculate the average weight for the $s$-th subcarrier.}	
		\State $\rW[4Ns:4Ns + 3N, 4Ns:4Ns + 3N] = w_s\mathbf{I}_{3N}.$ \Comment{Update the block identity matrix.}
				\EndFor
				\State $\bar{\rW}_\mathrm{p} = \rW$. 
	\end{algorithmic}
\end{algorithm} 
\par In Algorithm \ref{alg:algorithm2}, we take two steps to control an \emph{identical} weight applied for the measurements acquired through the same subcarrier based on the structure of the measurement illustrated in Fig. \ref{fig:wideband}: 1) diagonalizing the matrix to apply scaling to the measurements, 2) averaging the weights of the diagonal values corresponding to the same subcarrier. The updated weight $\bar{\rW}_\mathrm{p}$ is then used to calculate the forward propagation loss and update the weights $\rW_\mathrm{p}$, similar to the procedures taken in Algorithm \ref{alg:algorithm1}. 
\section{Simulations}\label{sec:simulationlast}
\par We consider a vehicular communication scenario in which the RSU is placed at a height of $5 \, \mathrm{m}$. The transmitter at the RSU is equipped with a $16 \times 16 $ UPA with phase shifters, i.e., $N=16$. We use phase shifters with low resolution,  with resolution as $q = 1$, $q=3$. The carrier frequency is set as $28\, \mathrm{GHz}$ and the operating bandwidth is $100\, \mathrm{MHz}$. In our simulation scenario, vehicles move along two lanes that are at a distance of $4\, \mathrm{m}$ and $7 \, \mathrm{m}$ from the foot of the RSU. We use Wireless Insite \cite{wireless_insite}, a ray tracing simulator, to obtain channel matrices between the transmitter and the receivers which are mounted on top of the vehicles. The beamspace prior corresponding to this scenario is shown in Fig. \ref{fig:beamspace_prior}. Note that our dataset contains both LOS and NLOS channels. The NLOS scenario  occurs when the link between the RSU and a vehicle is blocked by tall trucks.
\subsection{Results for a narrowband system}
\par The convolutional layer in our network consists of a single filter of dimensions $16 \times 16 \times 2$. This filter is used to optimize the base matrix in 2D-CCS. As shown in Fig. \ref{fig:2dccs_illus}, the last dimension of depth 2 is used to emulate the convolution using complex values. To emulate $M$ number of 2D-CCS-based channel measurements, the convolved output is subsampled to a $2M \times 1$ feature vector. The subsampled measurement vector is then fed into a cascade of four fully-connected layers with output dimensions of $80$, $256$, $512$ and $256$. For the beam alignment problem, a reasonable network is one that is invariant to the scaling of the channel. 
The bias at all the layers is forced to zero to ensure that the network satisfies the scale invariance property. We use around $20$k channels to train the network and $5$k other channels for testing. As explained in Section \ref{sec:scaling}, every channel in the training set is scaled so that $\Vert \mathbf{H} \Vert_{\mathrm{F}}=N$. For the test set, we use a common scaling for all the channels such that $\mathbb{E}[\Vert \mathbf{H} \Vert^2_{\mathrm{F}}]=N^2$. The channels are then restructured according to the procedure in Section \ref{sec:2dccs_implem}.
\begin{figure}[htbp]
\vspace{-6mm}
\centering
\includegraphics[trim = 2cm 0cm 0cm 0cm, clip = true, width = 7.0in]{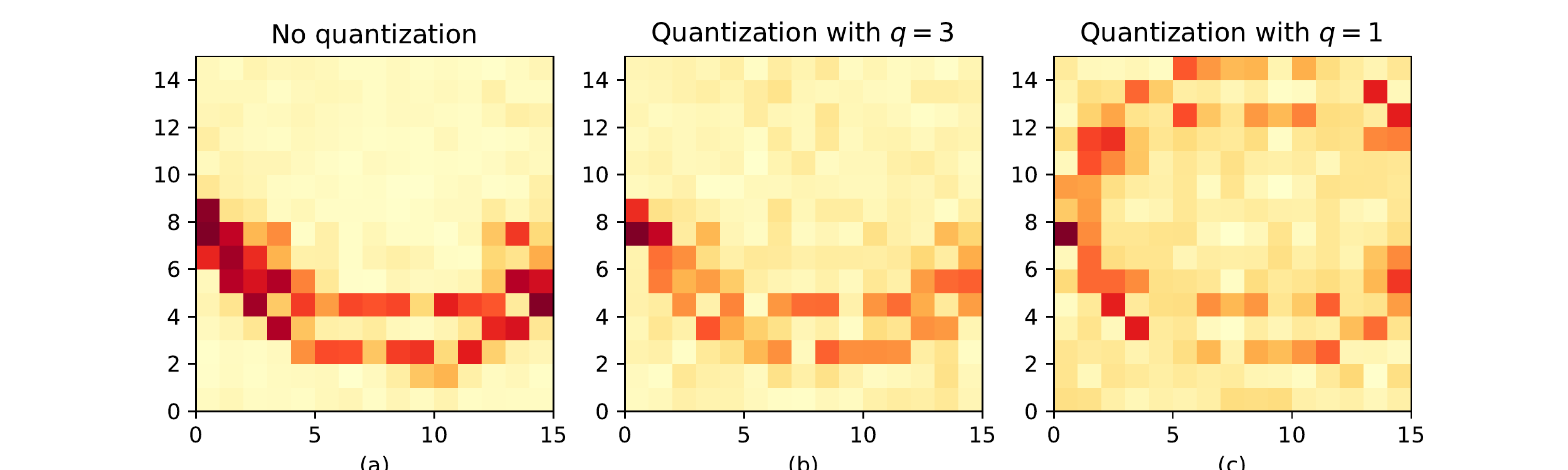}
\caption{ \small The plot shows the 2D-DFT magnitude of the base phase shift matrices optimized with our procedure for $M=40$. The mask optimized without quantization, i.e., $q=\infty$, based on the procedure in Section \ref{sec:2dccs_implem} is plotted in Fig. \ref{fig:mask}a. The masks optimized with few-bit phase shifters constraints are plotted in (b) and (c), which corresponds to the case of $q = 3$ in Fig. \ref{fig:mask}b, and $q = 1$ in Fig. \ref{fig:mask}c. Weight quantization slightly perturbs the beam pattern. The beams of phase shifters with different quantization resolution in the optimized 2D-CCS radiate power along the directions that are more likely to be optimal in the 2D-DFT beamspace. 
\normalsize}\label{fig:mask}
\end{figure}
\par We explain how our network is trained and provide insights into the optimized base matrix. For $M$ channel measurements, our method first samples $M$ distinct 2D-integer coordinates from an $N \times N$ grid at random and constructs the subsampling set $\Omega$. The $N \times N$ integer grid corresponds to the support of $\mathbf{G}$ in \eqref{eq:defn_G}. The $2M \times 1$ compressed channel measurement vector (including the real and imaginary part of the measurements) in 2D-CCS is perturbed by AWGN of variance $\sigma^2/2$. The SNR observed at the receiver when the transmitter uses a quasi-omnidirectional pattern is defined as $1/\sigma^2$. The proposed deep neural network is trained for $300$ epochs with the restructured channels and the associated class labels from the training set. A good base matrix is one whose beam pattern has a larger magnitude along the directions that are more likely to be optimal. A discrete version of this beam pattern is the 2D-DFT of the base matrix. 

In Fig. \ref{fig:mask}, we plot the masks optimized for different phase shifter resolutions. For the scenario corresponding to the beam prior in Fig. \ref{fig:beamspace_prior}, the 2D-DFT of the optimized base matrix with infinite-resolution phase shifters following the procedures in Section \ref{sec:2dccs_implem} after training is shown in Fig. \ref{fig:mask}a. Furthermore, the 2D-DFT magnitude of the base matrix with quantization derived based on Section \ref{sec:2dccs_quant_implem} are illustrated in Fig. \ref{fig:mask}b - c, which correspond to the cases of different quantization resolutions, $q = 3$ and $q = 1$. It can be observed that the beam pattern optimized with the given dataset captures the underlying statistical patterns of the channels and the resulting mask is well matched to the beamspace prior in Fig. \ref{fig:beamspace_prior}. It can also be observed phase shift matrix optimized with projected gradient descent are perturbations of the mask without phase shifter quantization shown in Fig. \ref{fig:mask}. Despite the perturbations, it can be observed that quantized masks also capture the important beam directions that are likely to be optimal corresponding to the beamspace prior in Fig. \ref{fig:beamspace_prior}.

We use two metrics for performance evaluation. As we focus on beam alignment using the 2D-DFT codebook, the SNR after beamforming can be expressed in terms of the 2D-DFT of the channel, i.e., $\mathbf{X}$. The SNR at the receiver when the transmitter applies the $(i,j)^{\mathrm{th}}$ element of the 2D-DFT codebook is $\mathrm{SNR}_{\mathrm{BF}}=|X(i,j)|^2/\sigma^2$. Alignment probability is defined as the fraction of channel realizations where the optimal beam pair in the 2D-DFT beamspace is predicted successfully by our proposed approach. The beamforming loss is defined as the average loss of the achieved beam reference signal received power (RSRP) compared to the maximum beam power with the optimal 2D-DFT beam aligned. 

\begin{figure}[!htb]
		\centering
		\includegraphics[width=5.0in ]{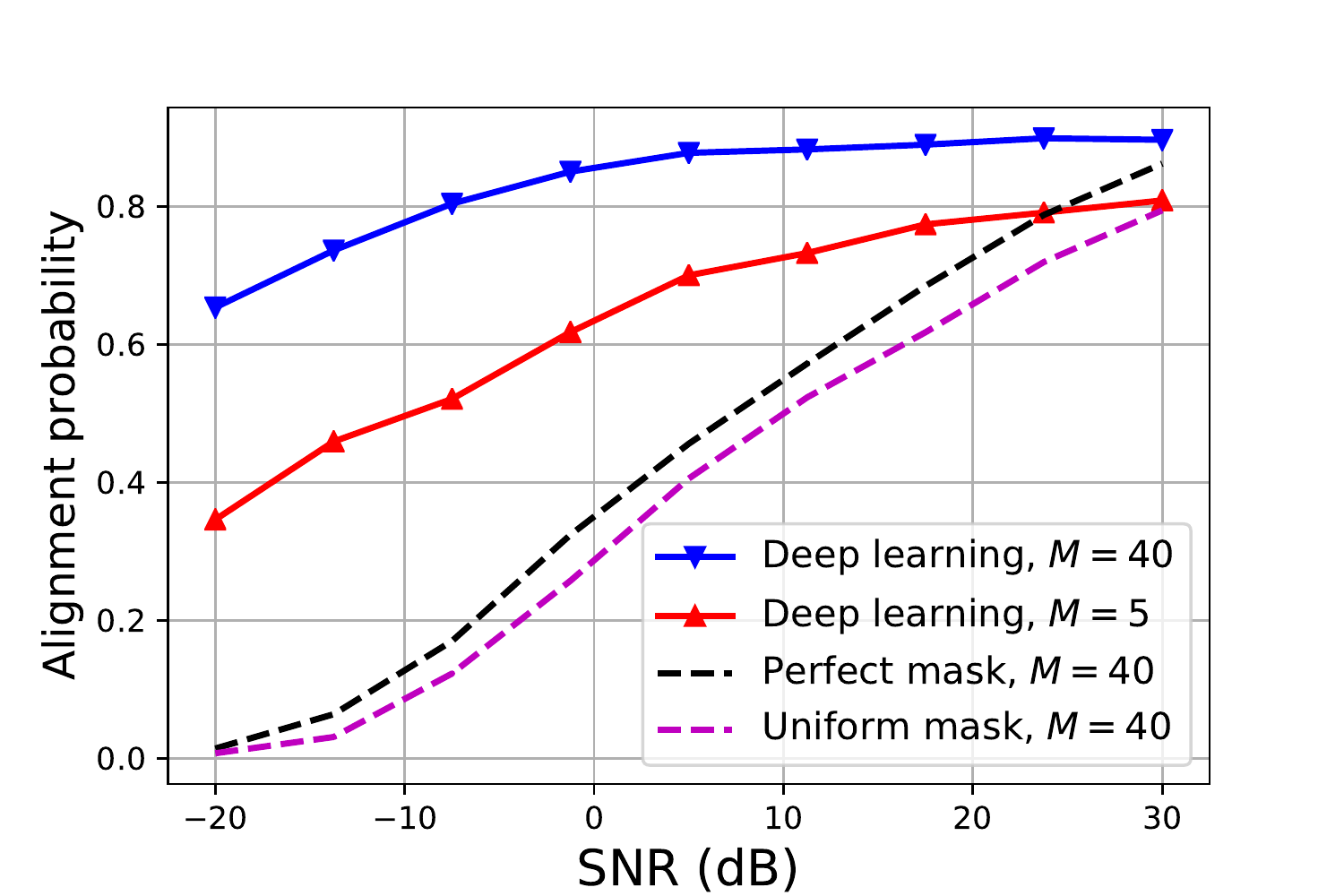}
		\caption{A comparison of the alignment probability achieved using the proposed deep-learning approach, the theoretical solution with perfect mask and uniform mask in \cite{wang2020sitespecific}. The solid curves represent the performance for the deep learning-based approach using different number of measurements $M = 40$ and $M = 5$. The dashed curves represent the theoretical solution with $M = 40$. }\label{fig:alignment}
			\end{figure}

\begin{figure}
					\centering
		\includegraphics[width=5.0in]{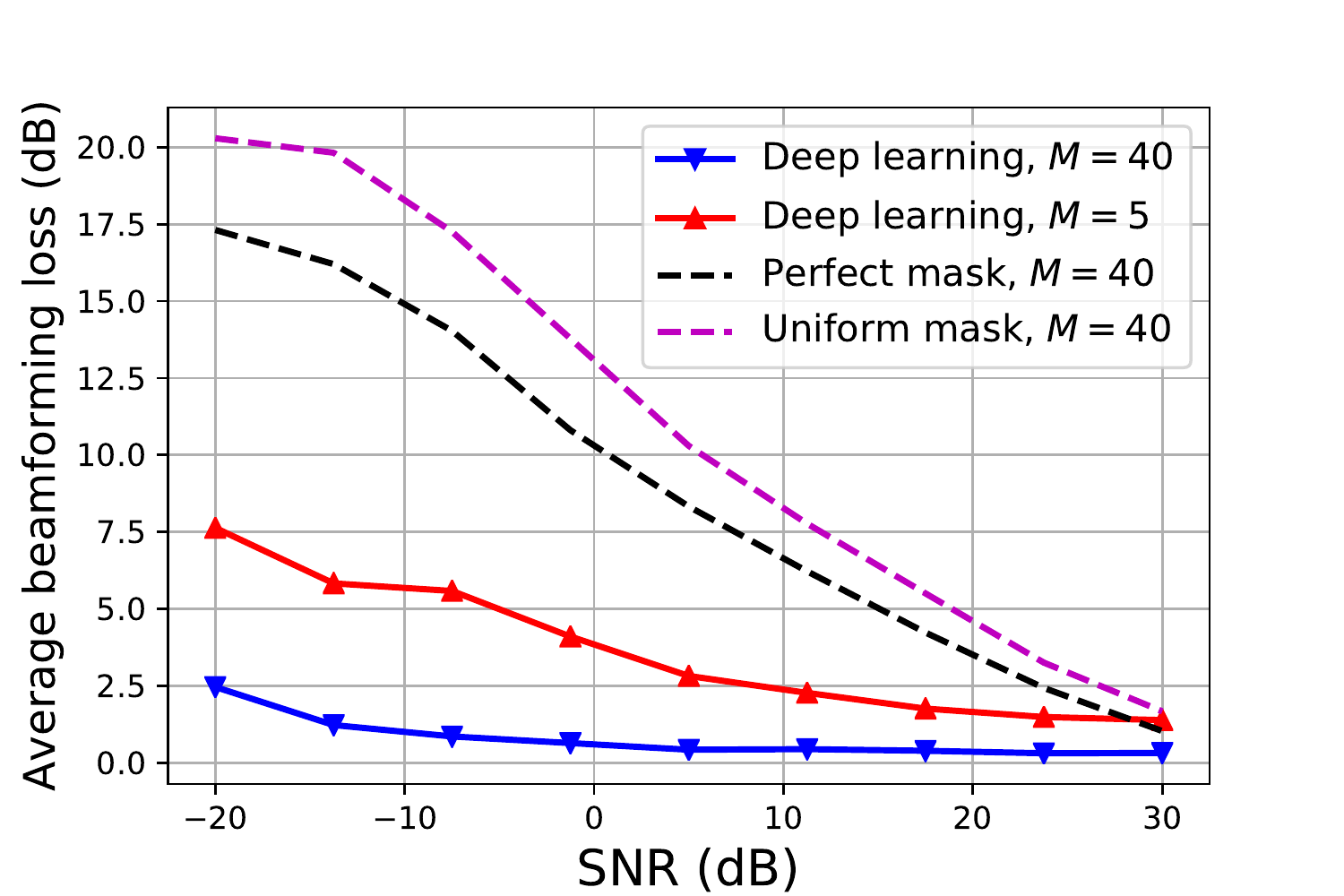}
		\caption{A comparison of the beamforming loss achieved compared to exhaustive beam search using the proposed deep-learning approach, the theoretical solution with perfect mask and uniform mask.}\label{fig:bfloss}
	\end{figure}

\begin{figure}
		\centering
		\includegraphics[width=5.0in]{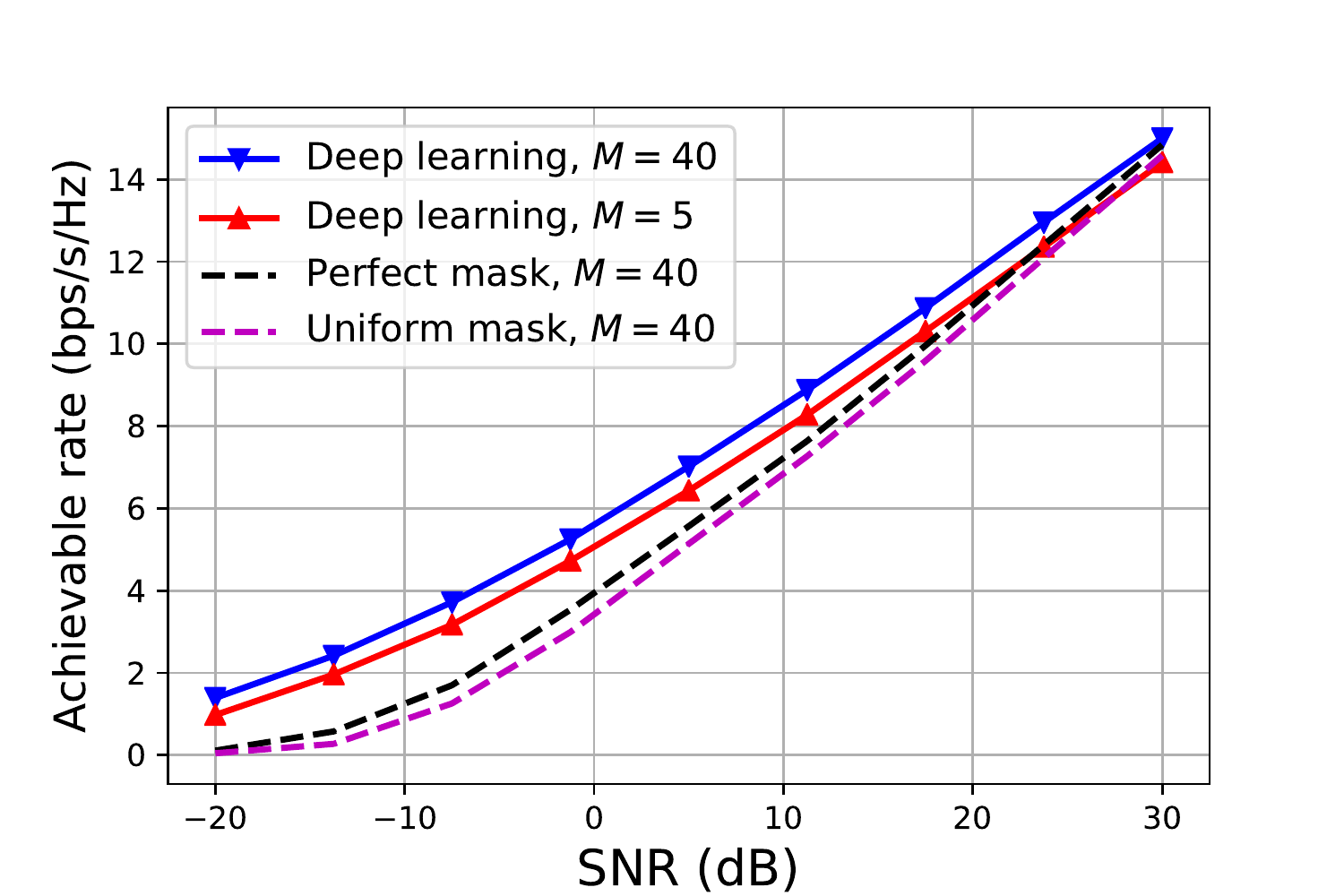}
		\caption{A comparison of the rates achieved using the proposed deep-learning approach, the theoretical solution with perfect mask and uniform mask.}\label{fig:rate}
	\end{figure}

In Figs. \ref{fig:alignment} - \ref{fig:rate}, we evaluate the alignment probability, beamforming loss, and rate achieved using the proposed deep learning-based beam alignment solution without hardware constraints, i.e., the phase shifters are infinite-resolution. We compare the deep learning-based approach with the theoretical solution derived in \cite{wang2020sitespecific}, where masks are optimized using convex optimization and beams are predicted based on an orthogonal matching pursuit (OMP) solution. It can be observed from Figs. \ref{fig:alignment} - \ref{fig:rate} that the deep learning-based approach achieves a large performance gain compared to theoretically derived approach in \cite{wang2020sitespecific}, especially at lower SNRs. Furthermore, it is shown that deep learning-based beam alignment solution is less susceptible to the changes of SNRs. Even at very low SNRs, it is still able to achievable reasonable alignment probability $\sim65\%$ if using $M = 40$ measurements and beamforming loss that is lower than $3$ dB. Last, it can be concluded from Figs. \ref{fig:alignment} - \ref{fig:rate} that the proposed approach achieves good performances using few measurements. Using $M = 5$ measurements achieves better performance than the theoretical solution with perfect mask using $M = 40$ measurements, when SNR is lower than $\sim 25$ dB. To conclude, the proposed beam alignment solution using deep learning is more robust to noise, and can achieve good alignment performance using a significantly reduced number of measurements. 

The reason that deep learning outperforms the theoretical approach in \cite{wang2020sitespecific} is two-fold: 1) the theoretical approach in \cite{wang2020sitespecific} leverages only the underlying channel angular domain distribution in a site-specific area, but our approach learns not only the angular distribution, reflected on the mask learned in Fig. \ref{fig:mask}, but also the distribution of the channel matrix $\rH$ itself. The full channel state information that is fed into the neural networks provides more complete information about the structure of the channel, and therefore largely improves the performance compared to the approach using only channel angular statistics in \cite{wang2020sitespecific}; 2) The formulation of \emph{perfect mask} in \cite{wang2020sitespecific}, is based on ideal assumptions that the channels are one-sparse, and theoretically-proved \emph{optimal mask} is not available yet. The mask learned using deep learning may introduce noise that is spread across the beamspace compared to the ideal mask that is optimized using convex optimization. The fully-connected layers following the 2D circular convolution, however, can fix the imperfections of the learned mask, thanks to the two-stage learning procedure in Section \ref{sec:scaling} that separates mask learning and beam alignment optimization.

\begin{figure}
		\centering
		\includegraphics[width=5.0in]{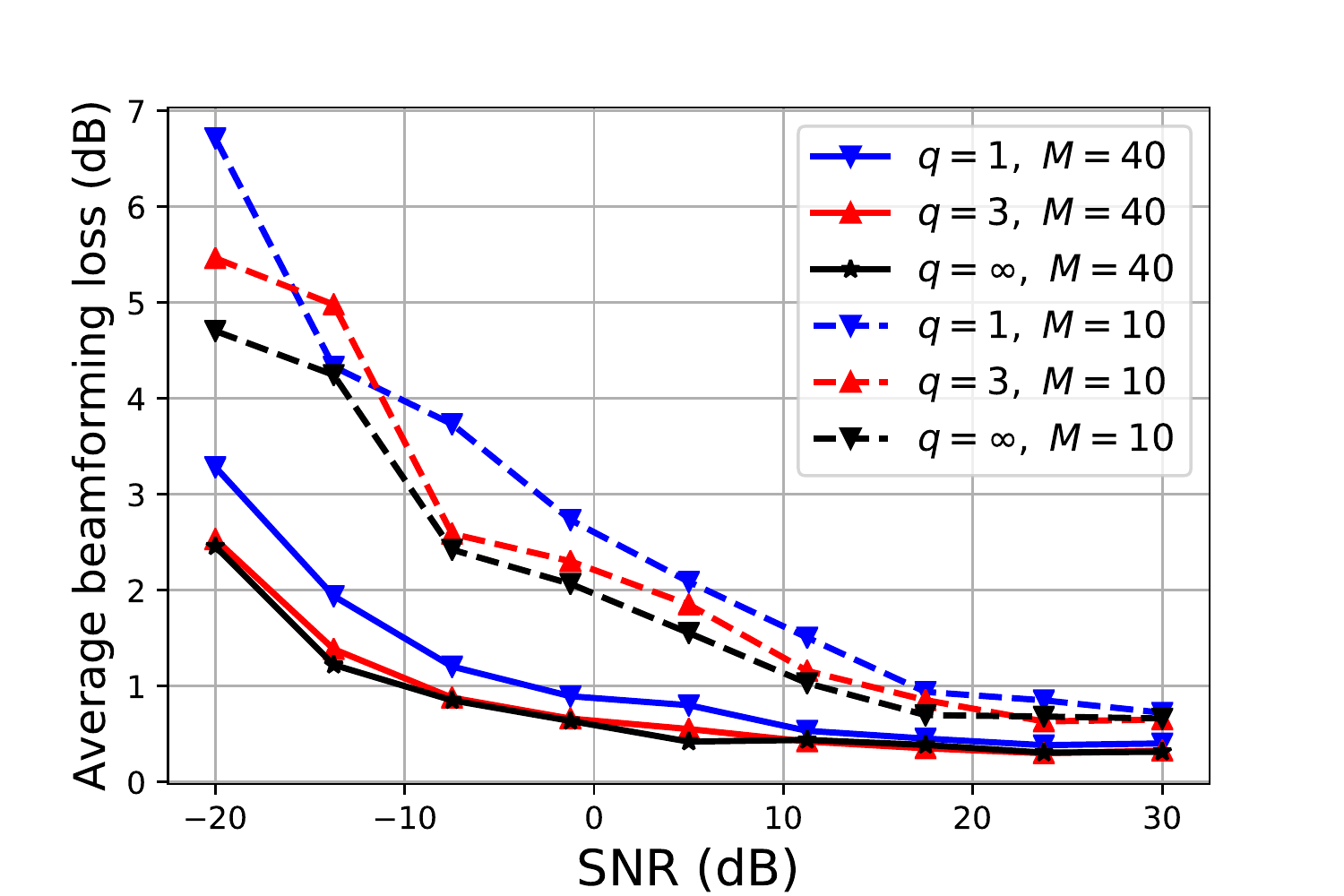}
		\caption{A	comparison of the average beamforming loss using different number of measurements under different quantization resolutions. The dashed curves represent the beamforming loss using $M = 10$ measurements. The solid curves represent the beamforming loss with $M = 40$. The black curves plot the performance without quantization, i.e., $q = \infty$. The red and blue curves plot the case with $q = 3$ and $q = 1$, respectively. }\label{fig:quant}
	\end{figure}

In. Fig \ref{fig:quant}, we evaluate the performance of the proposed framework using the projected gradient descent-based beam alignment optimization in Section \ref{sec:2dccs_quant_implem}, with different resolutions of phase shifters. We consider the cases where $M = 40$ and $M = 10$ measurements are used for beam alignment for each channel realization.  As can be observed from Fig. \ref{fig:mask}, the quantization of the phase shifters introduces perturbations to the mask, especially in the case of one-bit phase shifters. For $q = 1$, the approximately symmetrical beam pattern optimized due to the 1-bit phase shift constraint allocates measurement power to the beam directions that are not likely to be optimal, such as the \emph{upper half} of the beamspace in Fig. \ref{fig:beamspace_prior}. The difference between beam alignment performance with or without quantization is negligible almost across the whole SNR regime. This result is due to the additional fully-connected layers after the 2D convolutional layer, which predicts the optimal beam using the measurements acquired finite-resolution phase shifters. Though the beam pattern of the ideal phase shift matrix is perturbed a lot due to quantization, the subsequent fully-connected layers are able to adapt its weights and guarantee an optimized beam alignment solution for our problem. 

  	\begin{figure}
					\centering
		\includegraphics[width=5.0in]{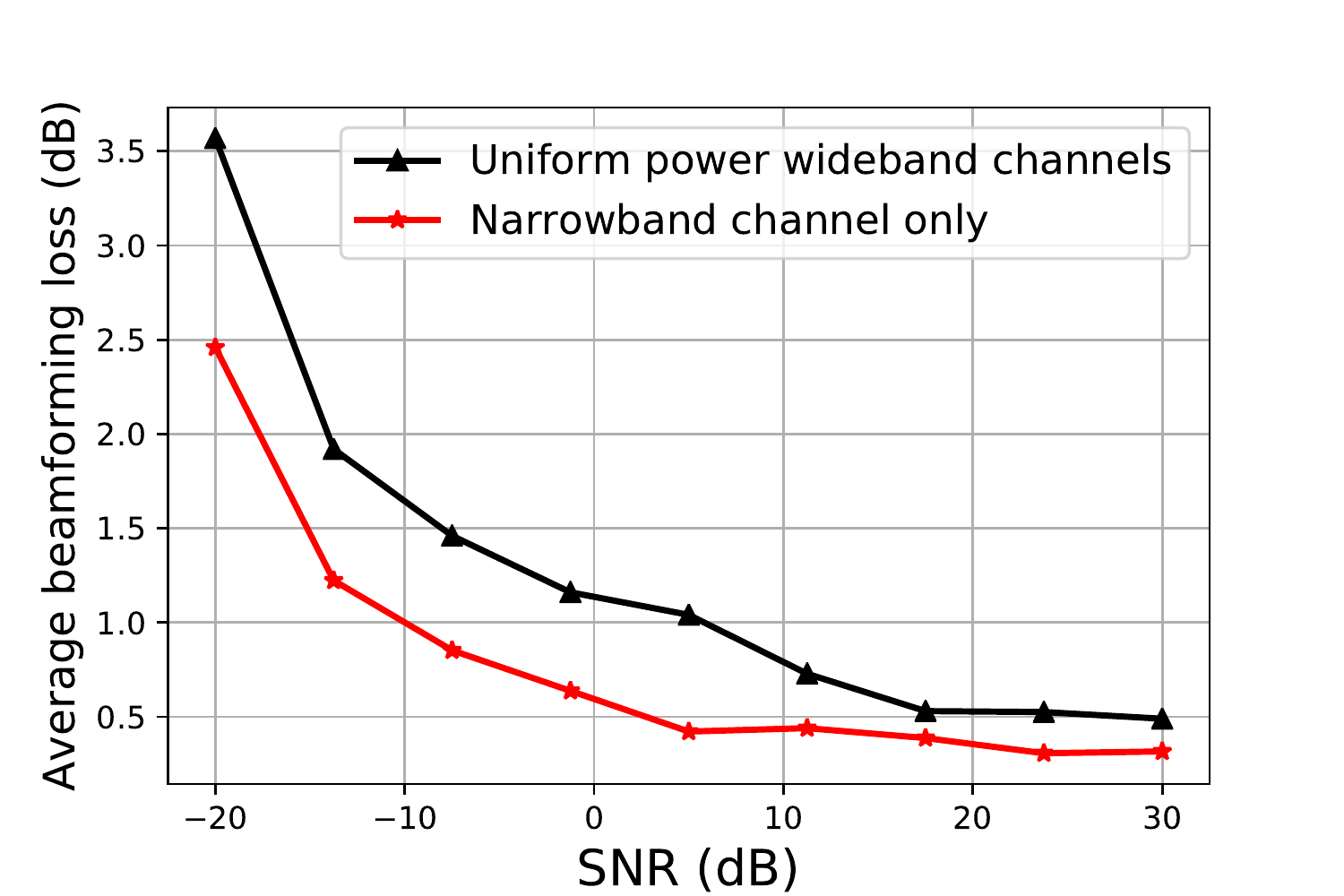}
		\caption{A comparison of the average beamforming loss using narrowband channels and wideband channels where power is uniformly allocated to different subcarriers. The black curve with markers $\triangle$ represents the wideband case, and the red curve with markers $\star$  represents the narrowband case.}\label{fig:wideband}
	\end{figure}

  	\begin{figure}
					\centering
		\includegraphics[width=5.0in]{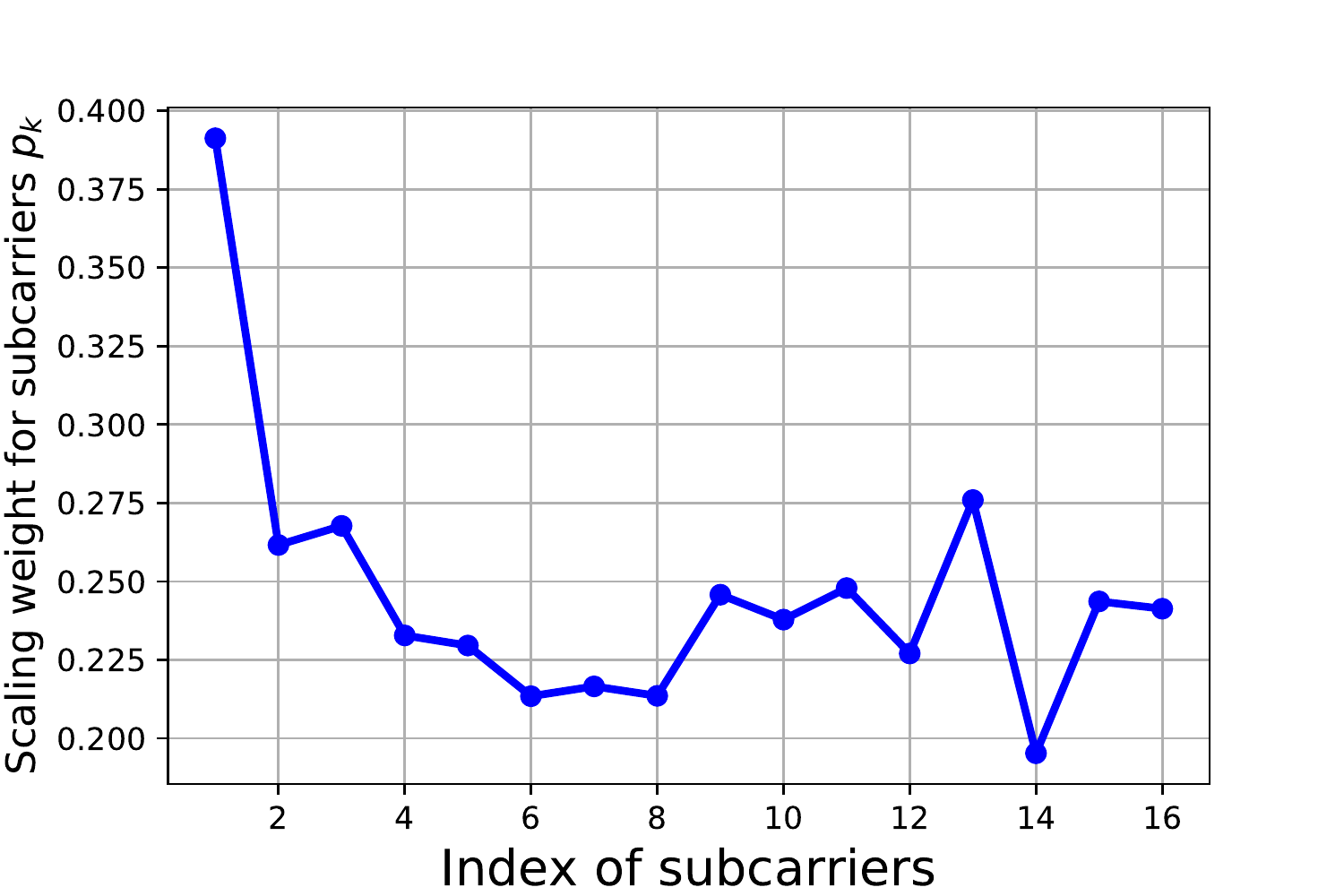}
		\caption{An illustration of the weight $p_k$ optimized for each subcarrier in Section \ref{sec:pow_alloc}, and it satisfies $\sum_{k\in\mathcal{S}} p_k^2 = 1$. }\label{fig:weight}
	\end{figure}

\subsection{Results for a wideband system}
In this section, we discuss the performance of the framework when extending the model input to wideband channels. As demonstrated in Section \ref{sec:wideband}, we add an additional linear layer layer  the identical subsampling to apply amplitude scaling to different subcarriers. We disable the bias in the linear layer, and diagonalize the optimization weight matrix in each weight update based on Algorithm \ref{alg:algorithm2}. 
In Fig. \ref{fig:wideband}, we compare the average beamforming loss achieved using the narrowband channels and the wideband channels with uniform power allocation to different subcarriers, as the input to the deep learning model. We evaluate the performance among different SNRs. Fig. \ref{fig:wideband} demonstrates advantage of simply applying narrowband channels in terms of the performance of beamforming loss. Furthermore, using narrowband channel as inputs means a model with largely reduced input dimension and fewer parameters that need to be optimized.  
In Fig. \ref{fig:weight}, we plot the result of the weight $p_s$ optimized for each subcarrier. From Fig. \ref{fig:weight}, we observe that the optimization layer tends to allocate more power to the first subcarrier, which gives $p_1 \approx 0.4$, while it allocates less power to other subcarriers, which are farther away from the DC subcarrier, where $p_k \approx 0.2, ~ \forall k\in\mathcal{S}\backslash\{1\}$. Furthermore, the multi-subcarrier channel input with weight optimization does not provide a significant performance gain compared to the narrowband case. Multi-subcarrier input with optimization gives similar beam alignment probability compare to the narrowband case. This result justifies the accuracy and efficiency of directly applying the first subcarrier in frequency-domain channels as inputs to conduct beam alignment.
\section{Conclusions}\label{sec:conclusionlast}
In this paper, we developed a novel approach for compressive beam alignment with mmWave phased arrays using deep learning. Our method is based on a structured compressed sensing technique called 2D-convolutional compressed sensing. In 2D-CCS, any CS matrix can be parameterized by a base matrix and a subsampling set. We demonstrated how to implement 2D-CCS using CNN. We showed that optimization based on deep learning leverages underlying statistics in the channels to design a phase shift matrix well-suited to the channel prior. Furthermore, we proposed a projected gradient descent-based method which incorporates low-resolution phase shifter constraint in the training. We defined a certain quantization function over the convolutional filters, i.e., the phase shift matrix, during the forward propagation. Lastly, we developed a beam alignment framework with wideband channel inputs. We designed a linear layer that optimizes amplitude scalings to different subcarriers for power allocation. 

We evaluated our proposed approach in a ray tracing-based dataset established in a typical urban vehicular network. We demonstrated superior alignment probability and negligible beamforming loss achieved using the proposed approach, compared to the solution that is theoretically derived in \cite{wang2020sitespecific}. We showed that low-resolution phase shift matrix brought perturbations to the optimized beam pattern, while such perturbations can be compensated by the following fully-connected layers. Furthermore, the results in the wideband power allocation revealed that multiple subcarriers cannot substantially improve the performance of mmWave beam alignment compared t simply using single subcarrier, i.e., the narrowband channel case in our paper. The first subcarrier in the wideband channel is sufficient to provide all necessary information for beam configuration. In our future work, we will develop low complexity techniques for beam prediction and also extend our approach to examine the performance of our approach in channel estimation. \bibliographystyle{IEEEtran}
\bibliography{refs}
\end{document}